\documentclass[12pt]{article}
\usepackage{graphicx}
\usepackage{natbib}
\usepackage{color}
\usepackage[fleqn]{amsmath}
\newcommand{\ddt}[1]{\frac{\mathrm{d}#1}{\mathrm{d}t}}
\newcommand{\ka}{ \ensuremath{{\mathrm{ka}}}}
\newcommand{\Ma}{ \ensuremath{{\mathrm{Ma}}}}
\newcommand{\figref}[1]{Figure~{\ref{#1}}}
\newcommand{\figrefp}[1]{(\figref{#1})}
\newcommand{\introduction}{\section{Introduction}}

\newcommand{\conclusions}{\section{Conclusion}}

\title{Why could ice ages be unpredictable? \\ {\small Submitted to Climate of the Past on the 4th February 2012. }}
\author{Michel Crucifix \\ michel.crucifix@uclouvain.be \\ 
\small {Earth and Life Institute, Georges Lemaitre Centre for Earth and Climate Research} \\ 
\small {Universit\'e catholique de Louvain, Louvain-la-Neuve, Belgium} }

\begin{document}
\graphicspath{{Figures/Pdf/}}
\let\oldinclude=\include \def\include#1{\oldinclude{Figures/Tex/#1}}
\let\oldinput=\input \def\input#1{\oldinput{Figures/Tex/#1}}
\nocite{lisiecki05lr04}
\maketitle

 \begin{abstract}
It is commonly accepted that the variations of Earth's orbit and obliquity control the timing of Pleistocene glacial-interglacial cycles. Evidence comes from power spectrum analysis of palaeoclimate records and from inspection of the timing of glacial and deglacial transitions. However, we do not know how tight this control is. Is it, for example, conceivable that random climatic fluctuations could cause a delay in deglaciation, bad enough to skip a full precession or obliquity cycle and subsequently modify the sequence of ice ages?   

To address this question, seven previously published conceptual models of ice ages are analysed by reference to the notion of generalised synchronisation. Insight is being gained by comparing the effects of the astronomical forcing with idealised forcings composed of only one or two periodic components.  
In general, the richness of the astronomical forcing allows for synchronisation over a wider range of parameters, compared to periodic forcing.  Hence, glacial cycles may conceivably have remained paced by the astronomical forcing throughout the Pleistocene. 

However, all the models examined here also show a range of parameters for which the structural stability of the ice age dynamics is  weak.  This means that 
small variations in parameters or random fluctuations may cause significant shifts in the succession of ice ages if the system were effectively in that parameter range. 
Whether or not the system has strong structural stability depends on the amplitude of the effects associated with the astronomical forcing, which significantly differ across the different models studied here. The possibility of synchronisation on eccentricity is also discussed and it is shown that a high Rayleigh number on eccentricity, as recently found in observations, is no guarantee of reliable synchronisation.

\end{abstract}
\introduction
\citet{hays76} showed that  southern ocean climate benthic records exhibit spectral peaks around 19, 23-24, 42 and 100 thousand years (thousand years are henceforth denoted `ka'). More or less concomitantly \citet{berger77} showed, based on celestial mechanics, that the power spectrum of climatic precession was dominated by periods of 19, 22 and 24~\ka, and that of obliquity was dominated by a period of 41~\ka. These authors concluded that the succession of ice ages is somehow controlled by the astronomical forcing. 

The much less cited paper by \citet{Birchfield78aa} is, however, at least as important. These authors considered a dynamical ice sheet model, which they forced by astronomically-induced variations in incoming solar radiation (insolation). They managed to reproduce grossly the spectral signature found by \citet{hays76}. However, subtle changes in the model parameters, well within the range allowed by physics, disturbed significantly the precise sequence of ice ages, without altering the power spectrum of ice volume variations. 

\begin{figure}[t]
\begin{center}
\includegraphics{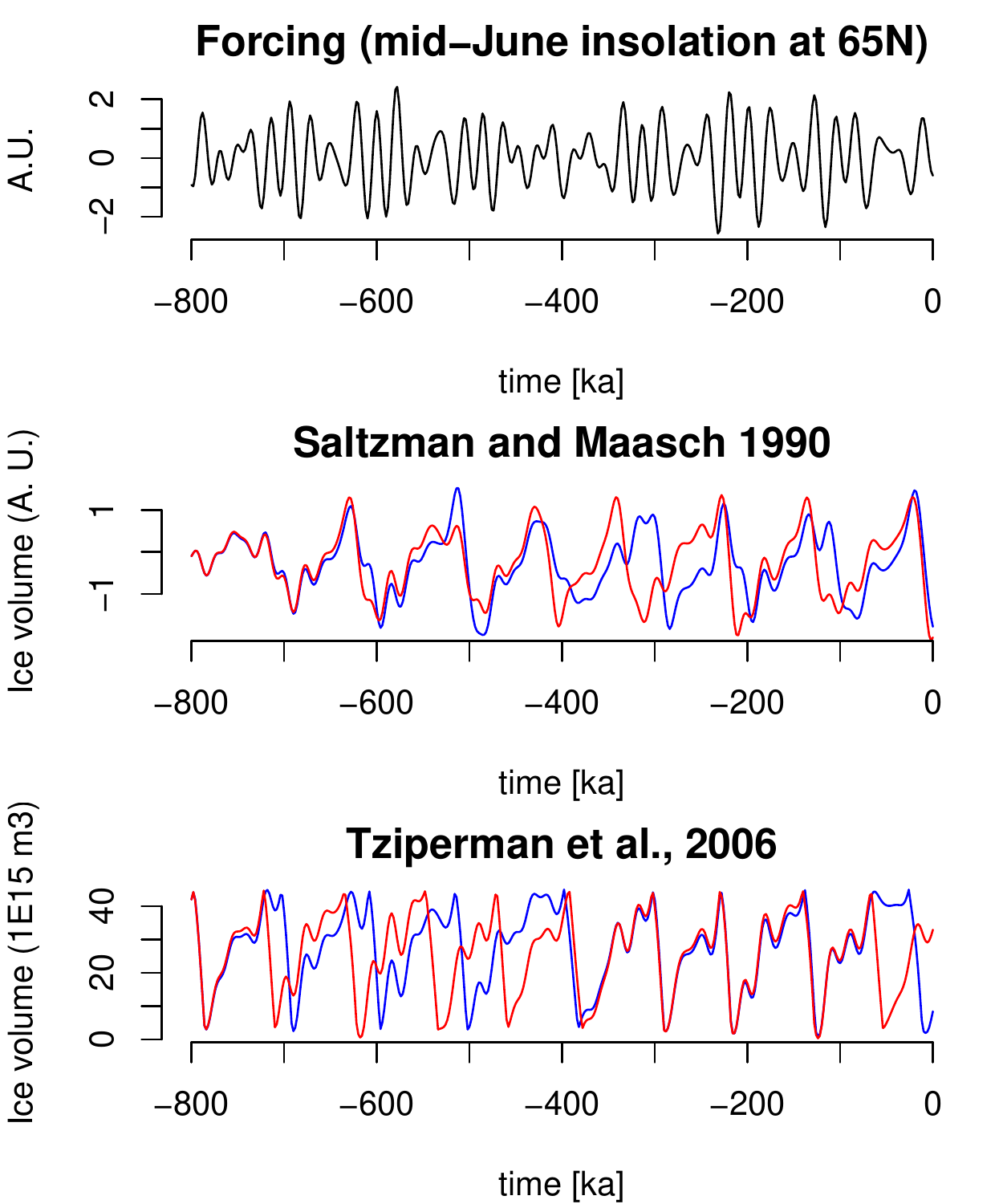}
\end{center}
\caption
{
 Ice volume simulated with two models previously published: \cite{saltzman90sm}  and \cite{tziperman06pacing}, forced by normalised insolation at 65$^\circ$ N. The blue lines are obtained with the published parameters; the latter were slightly changed to obtain the red ones: $p_0=0.262$~Sv instead of $0.260$~Sv in \cite{tziperman06pacing}, and $w=0.6$ instead of $w=0.5$ in \cite{saltzman90sm}. While the qualitative aspect of the curves are preserved, the timing of ice ages is affected by the parameter changes.  
}
\label{fig:figure1}
\end{figure}

It is this author's experience that some patient tuning is generally needed to reproduce the exact sequence of glacial-interglacial cycles with a conceptual model.
On \figref{fig:figure1} are shown two examples of ice volume history reproduced with models previously published in the palaeoclimate modelling literature \citep{saltzman91sm, tziperman06pacing}. In both cases, small changes in model parameters do, at some stage in the climate history, induce a shift in the sequence of ice ages. Sometimes this sensitivity is explicitly acknowledged by the authors \citep{paillard98, Imbrie11aa}, but not always, and this may have given the false impression that these models unambiguously confirm the tight control of astronomical forcing on ice ages.

Yet, as early as in 1980, \citet{imbrie80} posed the right questions. They wondered whether ``nonorbitally forced high-frequency fluctuations may have caused the system to flip or flop in an unpredictable fashion." They also noted that ``the regularity of the 100-ka  cycle,  and particularly its phase coherence with the 100-ka eccentricity cycle,  argue for predictability''. 

Let us comment these quotes:
\begin{description}
\item[Predictability:] The horizon of predictability of a system---i.e., the fact that one cannot predict its evolution arbitrarily far in time--- emerges as a combination of (1) our epistemic uncertainty on the system state, structure and its controlling environmental factors and (2) the stability of the system. All things being equal a stable system is more predictable than a chaotic one. What \citet{imbrie80} were asking is essentially how stable the climate system is with respect to non-astronomical fluctuations.
\item[Phase coherence with eccentricity:] The spectrum of eccentricity is dominated by a period of 413~ka, followed by four periods around 100~ka \cite[][Table 3]{berger78}. If the 100-ka eccentricity cycles have a strong controlling action on the succession of ice ages, then we expect the system to be quite stable to non-astronomical fluctations. In favour of this argument, \citet{Lisiecki10aa} recently documented a good coherence between the timing of eccentricity cycles and that of ice ages. 
\end{description}

Our purpose here is to understand the dynamical factors which may induce instability in the succession of ice ages. The approach is dynamics-oriented: we use tools from mathematics, and focus more on the understanding of the dynamics, than on the identification of physical mechanisms. 
Though, it will not be concluded whether or not glacial-interglacial cycles are indeed predictable or not. This requires an additional step of statistical inference, which is left for another article. 

Which model to use? There are many models of ice ages, spanning different orders of complexity and based on different physical interpretations. 
We will therefore work with different models, but only of the class of the simplest ones.  This choice offers us a greater flexibility in analysing model dynamics with computing intensive techniques, and it also allows us to keep our hypotheses to a minimum.

Indeed, most of the simplest models of ice ages \citep{saltzman90sm, saltzman91sm, paillard04eps,tziperman06pacing,Imbrie11aa} share a number of characteristics:
\begin{enumerate}
\item  These are dynamical systems: climate has a memory (in contrast to \citet{milankovitch41}); 
\item the astronomical forcing is introduced as an additive or quasi-additive forcing term, which involves a combination of precession and obliquity;
\item there are non-linear terms involved in the internal system dynamics, which induce episodically conditions of instability. The general hypothesis is that high glaciation levels are unstable. The instability conditions may lie implicitly in the system dynamics \citep{saltzman90sm},  or postulated explicitly by means of a threshold criteria \cite[as in ][]{paillard98, Imbrie11aa}. The threshold may be a function of precession and obliquity \citep{Parrenin12ab}, and a dependency on eccentricity was also proposed \citep{Rial04aa}. 
\end{enumerate}
System instability is an important aspect of Pleistocene theory. It is a convenient starting point to explain the existence of large climatic fluctuations such as deglaciations, even when the astronomical forcing is weak. Termination V, which occurred  400~ka ago, is an often-cite example \citep{paillard01rge}. 
Instability may also explain the emergence of 
100-ka  climatic cycles independently of the effect of eccentricity (see, e.g. \cite{Crucifix12aa} for a review).

The present article is structured as follows. The discussion starts with the van der Pol oscillator forced by astronomical forcing. As the other models cited so far, this is a dynamical system that combines the accumulative action of astronomical forcing with an instability mechanism causing regime changes. The \citet{vanderpol26} Pol oscillator was first introduced as a model of an electronic circuit and it has been studied for over 80 years. This gives us the possibility to anchor the present work in a long tradition of dynamical system theory. 
Next,  the analysis techniques used with the van der Pol oscillator are applied to 6 other models previously published in the literature. We will then  be able to determine which conclusions seem the most robust. 

\section{The van der Pol oscillator}
\subsection{Model definition}
The van der Pol model can be introduced as a dynamical system of two coupled ordinary differential equations:

\begin{equation}
\left\{  \parbox{\fill}{ \vskip-1em
\begin{eqnarray*}
\frac{\mathrm{d}x}{\mathrm{d}\,t} &=&  - \frac{1}{\tau} \left( F\left( t \right) + \beta + y \right). \\
\frac{\mathrm{d} y}{\mathrm{d}\,t} &=& \frac{\alpha}{\tau}( y - y^3/3 + x),
\end{eqnarray*} }
\right.
\label{eq:vdp} 
\end{equation}
with: $(x,y)$ the climate  state vector, $\tau$ a time constant, $\alpha$ a time-decoupling factor, $\beta$ a bifurcation parameter and $F(t)$ the forcing. The parameter $\beta$ does not appear in the original van der Pol equations. 
The present variant is sometimes referred to as the \textit{biased van der Pol model}.
 
The autonomous (i.e. $F(t)=0$) model displays self-sustained oscillations as long as  $|\beta|<1$. For later reference the period of the unforced oscillator is denoted $T_n(\tau)$. The variable $x$ follows then a saw-tooth periodic cycle, the asymmetry of which is controlled by $\beta$.

Variable $y$ varies the more abruptly that $\alpha$ is high. Here we chose $\alpha=30$, so that $y$ may be termed the `fast' variable. 
In climate terms, $x$ may be interpreted as a glaciation index, which accumulates slowly the effects of the astronomical forcing $F(t)$, while $y$, which shifts between approximately $-1$ and $1$, 
might be interpreted as some representation of the ocean dynamics. This is admittedly arguable and other interpretations would be possible. Keep in mind that the van der Pol model is only used here to identify phenomena emerging from the combination of limit cycle dynamics and astronomical forcing,  and models with better physical justifications are analysed section \ref{sect:others}.

\begin{figure}[t]
\includegraphics{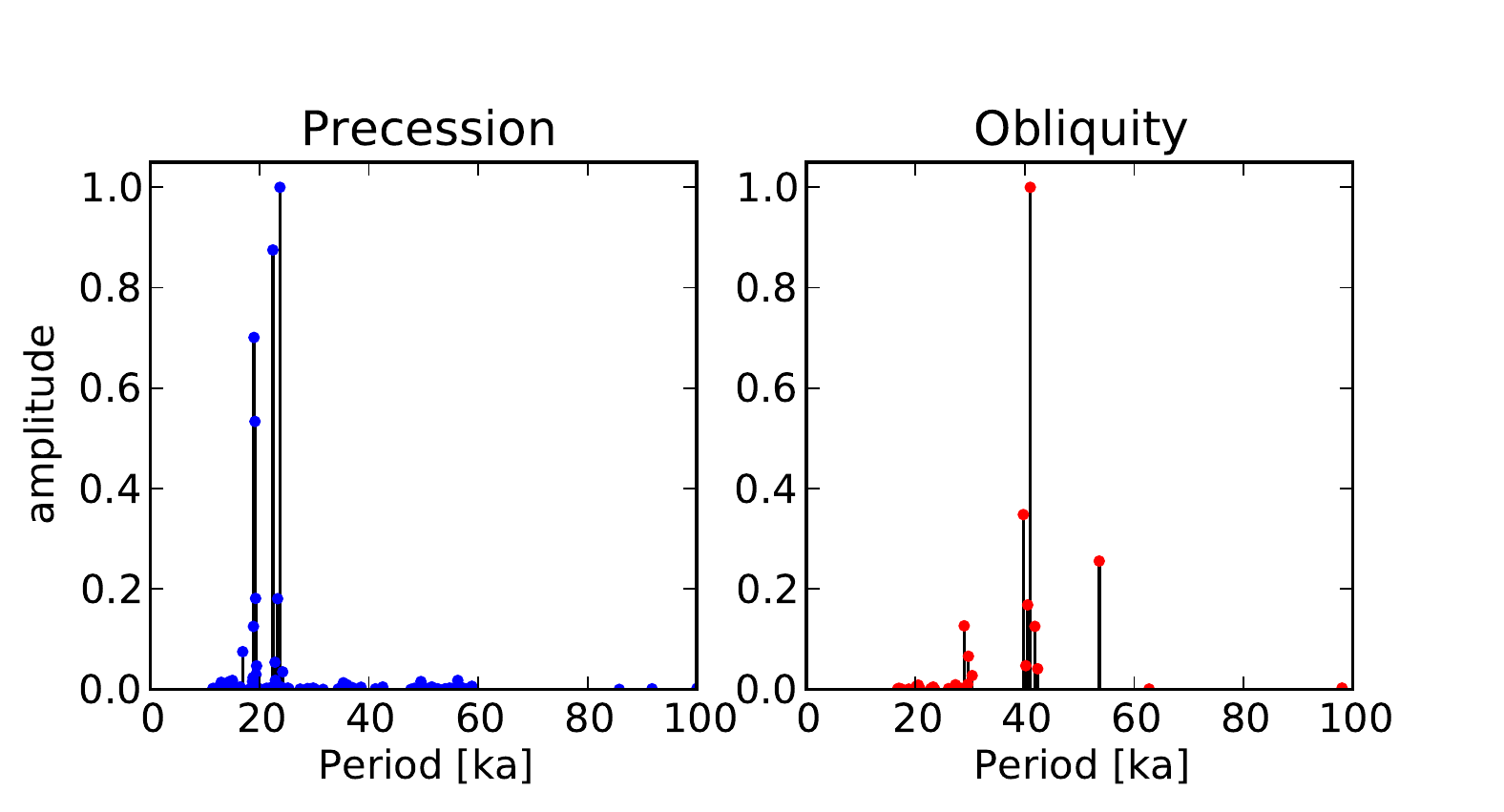}
\caption
{
 Spectral decomposition of precession and obliquity given by \cite{berger78}, scaled here such that the strongest components have amplitude 1. }
\label{fig:astro}
\end{figure}

\begin{figure}[t]
\begin{center}
\includegraphics[scale=1.0]{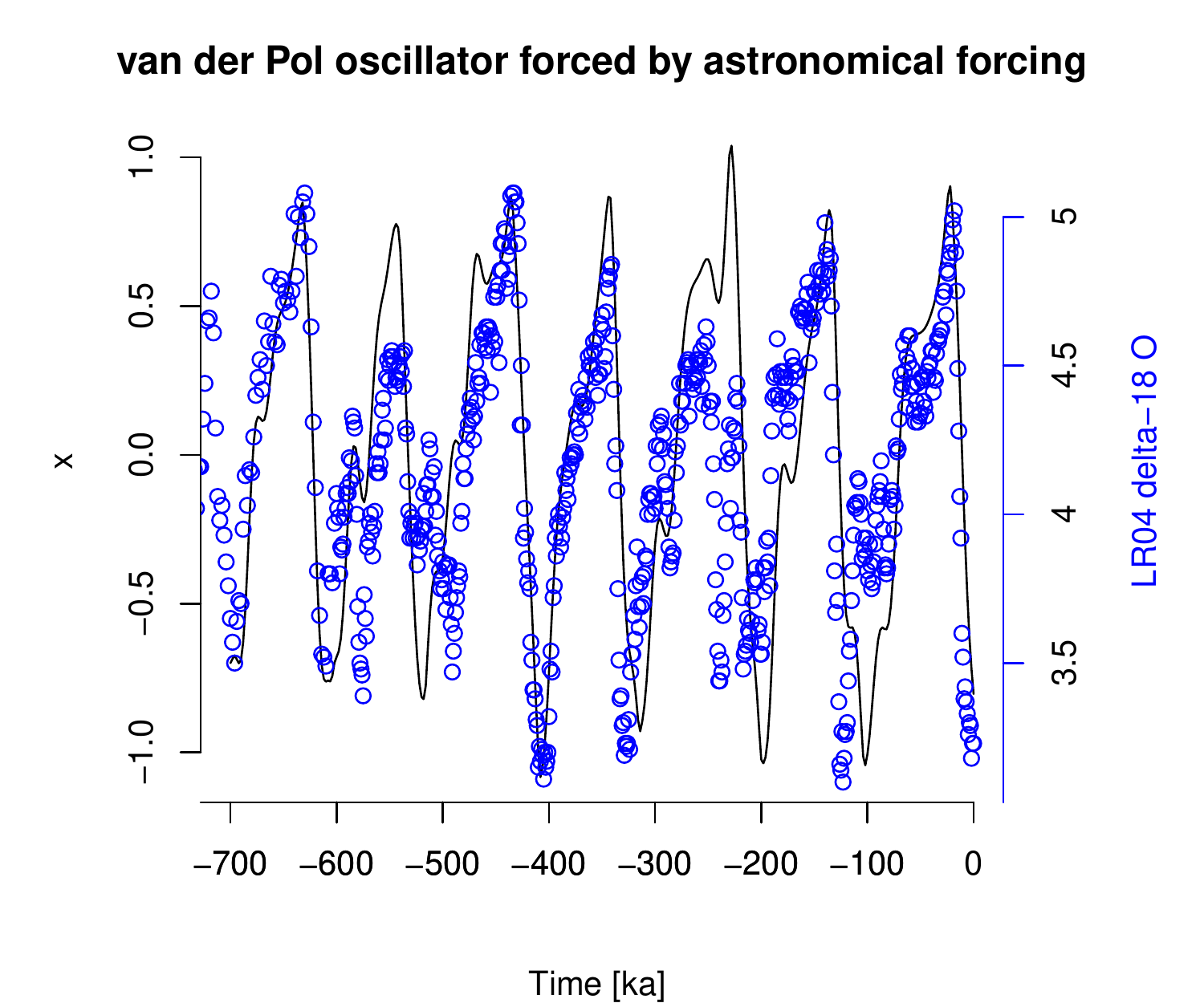}
\end{center}
\caption
{
Simulation with the van der Pol oscillator (eq.: \ref{eq:vdp}, \ref{eq:insol_vdp} and \ref{eq:insol_pi} -- \ref{eq:insol_eps}), forced by astronomical forcing and compared with the benthic record of \cite{lisiecki05lr04}. Parameters are : $\alpha=30$, $\beta=0.75$, $\gamma_p=\gamma_o=0.6$, $\tau=36.2\,\ka$. 
}
\label{fig:vdpfit}
\end{figure}

In ice age models the forcing function is generally one or several insolation curves, computed for specific  seasons and latitudes. The rationale behind this choice is that whichever insolation is used  it is, to a very good approximation, a linear combination of climatic precession and obliquity (\citet{Loutre93aa}, see also Appendix \ref{sect:appinsol}). The choice of one specific insolation curve 
may be viewed as a modelling decision about the effective forcing phase of climatic precession, and the relative amplitudes of the forcings due to precession and obliquity.
In turn, climatic precession and obliquity can be expressed as a sum of sines and cosines of various amplitudes and frequencies \citep{berger78}, so that  $F(t)$ can be modelled as a linear combination of about a dozen of dominant periodic signals, plus a series of smaller amplitude components. They are shown on  Figure \ref{fig:astro}. More details are given in section \ref{sect:fullastr}.
With these hypotheses the van der Pol model may be tuned so that the benthic curve over the last 800,000 years is reasonably well reproduced (\figref{fig:vdpfit}).

An abundant literature analyses the response of the van der Pol oscillator to a periodic forcing \cite[e.g.][and ref. therein]{Mettin93aa,Guckenheimer03aa}. The response of oscillators to the sum of two periodic forcings has been the focus of attention because it leads to the emergence of `strange non-chaotic attractors', on which we will come back \citep{Wiggins87aa, Romeiras87aa, Kapitaniak90ab, Kapitaniak93aa, Belogortsev92aa, Feudel97aa, Glendinning00aa}. To our knowledge, however, there is no systematic study of the response of an oscillator to a signal of the form of the astronomical forcing, except for preliminary work of our group \citep{De-Saedeleer12aa}. \cite{letreut83}, for example,  represented the astronomical forcing as a sum of only two or three periodic components and it will be shown  here that it matters to consider the astronomical forcing with all its complexity. 

\subsection{Periodic forcing}
Consider a sine-wave forcing ($F(t)=\gamma\sin(2\pi / {P_1} t + \phi_{P1})$), with period $P_1= 23,716$  years and $\phi_{P1}=32.01^\circ$. This is the first component of the harmonic development of  climatic precession \citep{berger78}. 
If certain conditions are met---they will soon be given---the van der Pol oscillator may become synchronised on the forcing. \textit{Synchronised} means, in this particular context, that the response of the system displays $p$ cycles within $q$ forcing periods, where $p$ and $q$ are integers. It is said that the system is in a $p:q$ synchronisation regime \cite[][p. 66-67]{Pikovski01aa}. The output is periodic, and its period is equal to $q \times P$.

\begin{figure}[t]
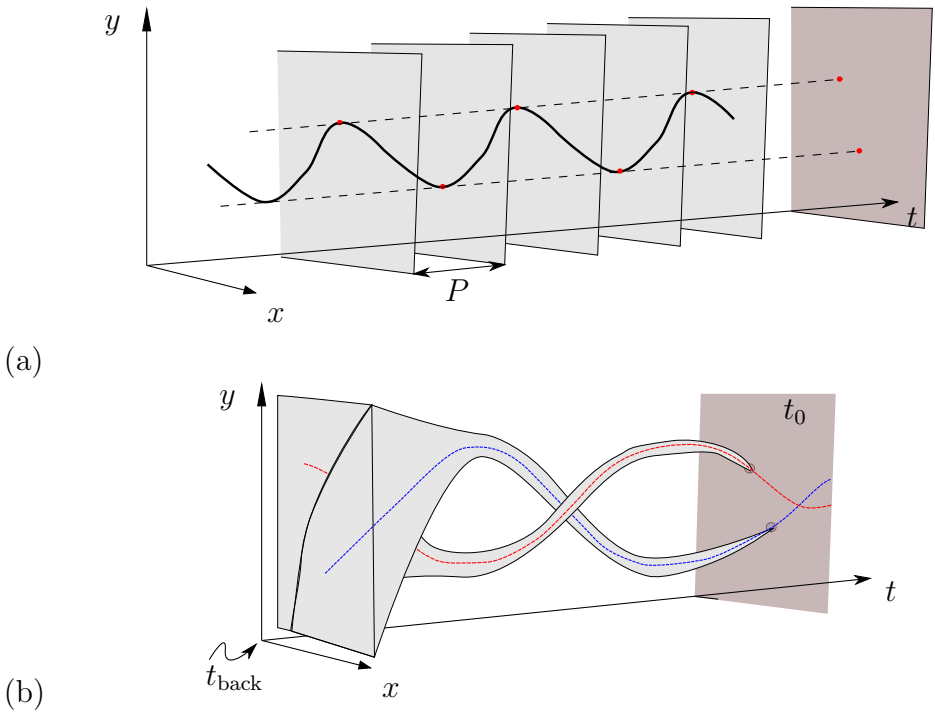

\begin{tabular}{ll}
(a) & \
\input{strob.pdf_tex} \\
(b) & 
\input{pullback.pdf_tex} \\
\end{tabular}
\caption{(a) The stroboscopic section is obtained by superimposing the system state every forcing period ($P$), here illustrated for a 2:1 synchronisation. (b) The pullback section at time $t_0$ is obtained by superimposing the system states obtained by initialising the system with the ensemble of all possible initial conditions far back in time, here at $t_{\mathrm{back}}$. The particular example shows a global pullback attractor made of two local pullback attractors (in dashed red and blue), the sections of which are seen at $t$. Also shown is the convergence of initial conditions towards the pullback attractors, in grey. }
\label{fig:pullback}
\end{figure}

There are several ways to identify the  synchronisation in the output of a dynamical system. One method is to plot the state of the system at a given time $t$, and then superimpose on that plot the state of the system at every time $t+nP$, where $n$ is integer. The system is synchronised  if only $q$ distinct points appear on the graph,  discarding transient effects associated with initial conditions. These correspond to the stable fixed points of the iteration bringing the system from $t$ to $t+qP$. In the following we refer to this kind of plot as a ``stroboscopic section" of period $P$ (\figref{fig:pullback}a). If the system is not synchronised, then there are two options: the stroboscopic section is a closed curve (the response is \textit{quasi-periodic}), or a figure with a strange geometry (the response is \textit{a-periodic}). 

There is another, equivalent way to identify synchronisation. Suppose that the system is started from arbitrary initial conditions. Then, plot the system state at a given time $t$, long enough after the initial conditions. Repeat the experiment with another set of initial conditions, superimpose the result to the plot, and so on with a very large number of different initial conditions. In doing so one constructs the section of the global pullback attractor at time $t$ (henceforth referred to as the \emph{pullback section}); the pullback attractor itself being the continuation of this figure over all times (\figref{fig:pullback}b). Each component of the global pullback attractor (two aro illustrated on \figref{fig:pullback}b) is a local pullback attractor.  
\citet[chap. 2]{Rasmussen00aa} reviews all the relevant mathematical formalism. 

\begin{figure}[t]
\includegraphics{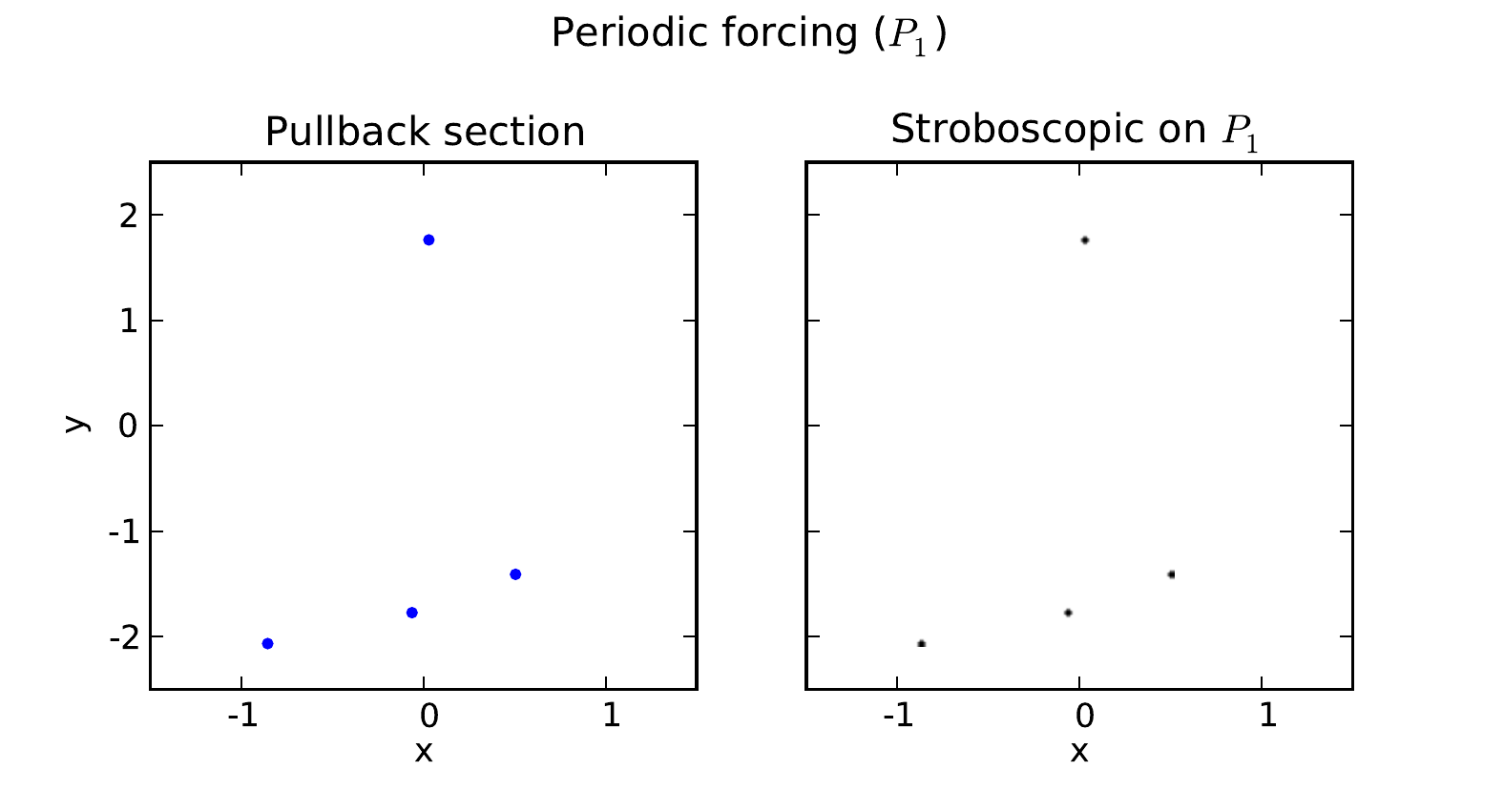}
\caption
{
Pullback (at $t_0=0$) and stroboscopic sections ($t=0 + nP_1$) obtained  with the van der Pol oscillator, with parameters $\alpha=30$, $\beta=0.7$, $\tau=36$ forced by $F(t)=\gamma \sin(2\pi/P_1 t + \phi_{P1})$, $P_1=23.7\,\ka$, $\phi_{P1}=32.01^\circ$ and $\gamma=0.6$. The two plots indicate a case of $4:1$ synchronisation, and they are identical because the forcing is periodic.
}
\label{fig:strob_periodic}
\end{figure}

In the particular case of a periodic forcing, the stroboscopic section and the pullback section are often identical  \figrefp{fig:strob_periodic}.
\footnote{
This property derives from the system invariance with respect a time translation by $P$. 
There  will be, however, cases where different initial conditions will create different stroboscopic plots. For example two $1:2$ synchronisation regimes co-exist in the forced  van der pol oscillator, so that there are four distinct local pullback attractors, while only two points will appear on a stroboscopic plot started from a single set of initial conditions. Rigorously, the global pullback attractor at time $t$ is identical to the global attractor of the iteration $t+nP$, and the global pullback attractor at two times $t$ and $t^\prime$ are homeomorphic.}

\begin{figure}[t]
\begin{center}
\includegraphics{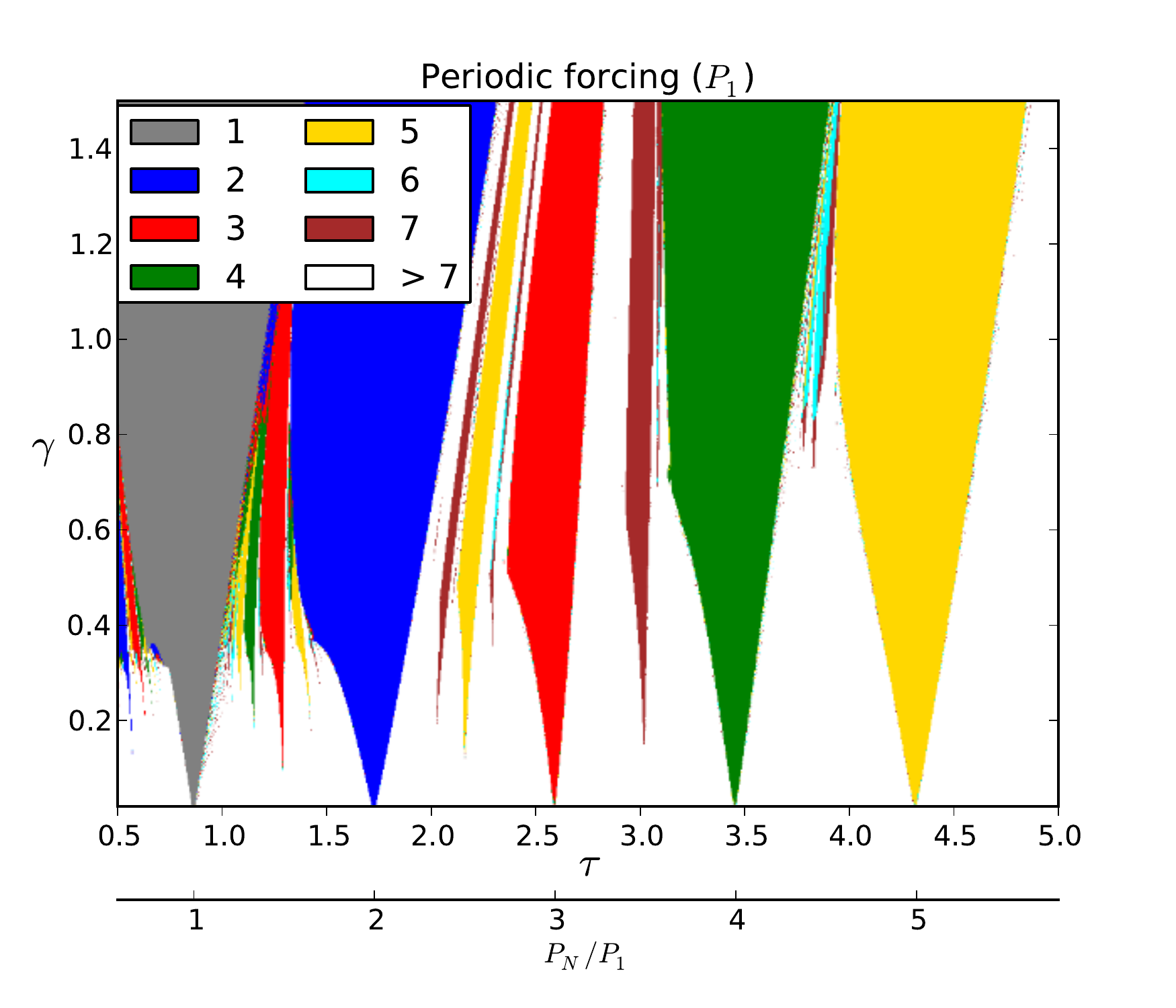}
\end{center}
\vskip-2.5em
\caption
{
Bifurcation diagram obtained by counting the number of points on the pullback section in the van der Pol oscillator ($\alpha=30, \quad \beta=0.7$) and  $F(t)=\gamma \sin( 2\pi / P_1 t + \phi_{P1})$. The two x-axes indicate $\tau$ and the ratio between the natural system period and the forcing, respectively.  One observes the synchronisation regimes corresponding to 1:1, 2:1, 3:1, 4:1 and 5:1, respectively (gray, blue, red, green, yellow) and, intertwinned, higher order synchronisations including $2:3$, $2:5$, $3:2$ etc.  White areas are weak or no synchronisation. Graph constructed using $t_\mathrm{back}=-10$Ma (see Figure \ref{fig:convergence} and text for meaning and implications.)
}
\label{fig:card_periodic}
\end{figure}

The number of points on the pullback section may then be estimated for different combinations of parameters and we can use this as a criteria to detect synchronisation. This is done on \figref{fig:card_periodic} for a range of $\gamma$ and $\tau$. 
It turns out that synchronisation regimes are organised in the form of triangles, known in the dynamical system literature as \textit{Arnol'd tongues} \cite[][p. 52]{Pikovski01aa}. A  $p:q$ synchronisation regime appears when the ratio between the natural period ($T_n$) and the forcing period is \textit{close} to $q/p$. The tolerance, i.e., how distant this ratio can afford to be with respect to $q/p$, increases with the forcing amplitude and decreases with $p$ and $q$.
Synchronisation is weakest (least reliable) near the edge of the tongues. Unreliable synchronisation characterises a system that is synchronised, but in which small fluctuations may cause episodes of desynchronisation. In the particular case of periodic forcing the episode of desynchronisation is called a  phase slip, as is well explained in \citet[][p. 54]{Pikovski01aa}

Using the pullback attractor to identify synchronisation is not a very efficient method in the periodic forcing case. Arc-length continuation methods are faster and more accurate \citep[e.g.,][and ref. therein]{Schilder07aa}. It is shown however in \cite{De-Saedeleer12aa}  that the pullback section method gives results that are acceptable enough for our purpose, and it is adopted here because it provides a more intuitive starting point to characterise synchronisation with multi-periodic forcings.

\subsection{Synchronisation on two periods}
Consider now a forcing function that is the sum of two periodic signals. Two cases are considered here: the two forcing periods differ by a factor of about 2, and the two forcing periods are close.
\subsubsection{$P_1 = 23.716\ka$\  and $O_1 = 41.000\ka$}
We adopt $F(t)=\gamma [\sin (2\pi/P_1 t + \phi_{P1})) + \cos (2\pi / O_1 t + \phi_{O1})]$, with $P_1 = 23.716\ka$, $O_1 = 41.000\ka$ and $\phi_{P1}=32.01^\circ$ and $\phi_{O1}=251.09^\circ$. $P_1$ is the first period in the development of precession,  $O_1$ is the first period in the development of obliquity and $\phi_{P1}$,$\phi_{O1}$ the corresponding phases given by \cite{berger78}, so that $F(t)$ may already be viewed as a very rough representation of the astronomical forcing.

\begin{figure}[t]
\includegraphics{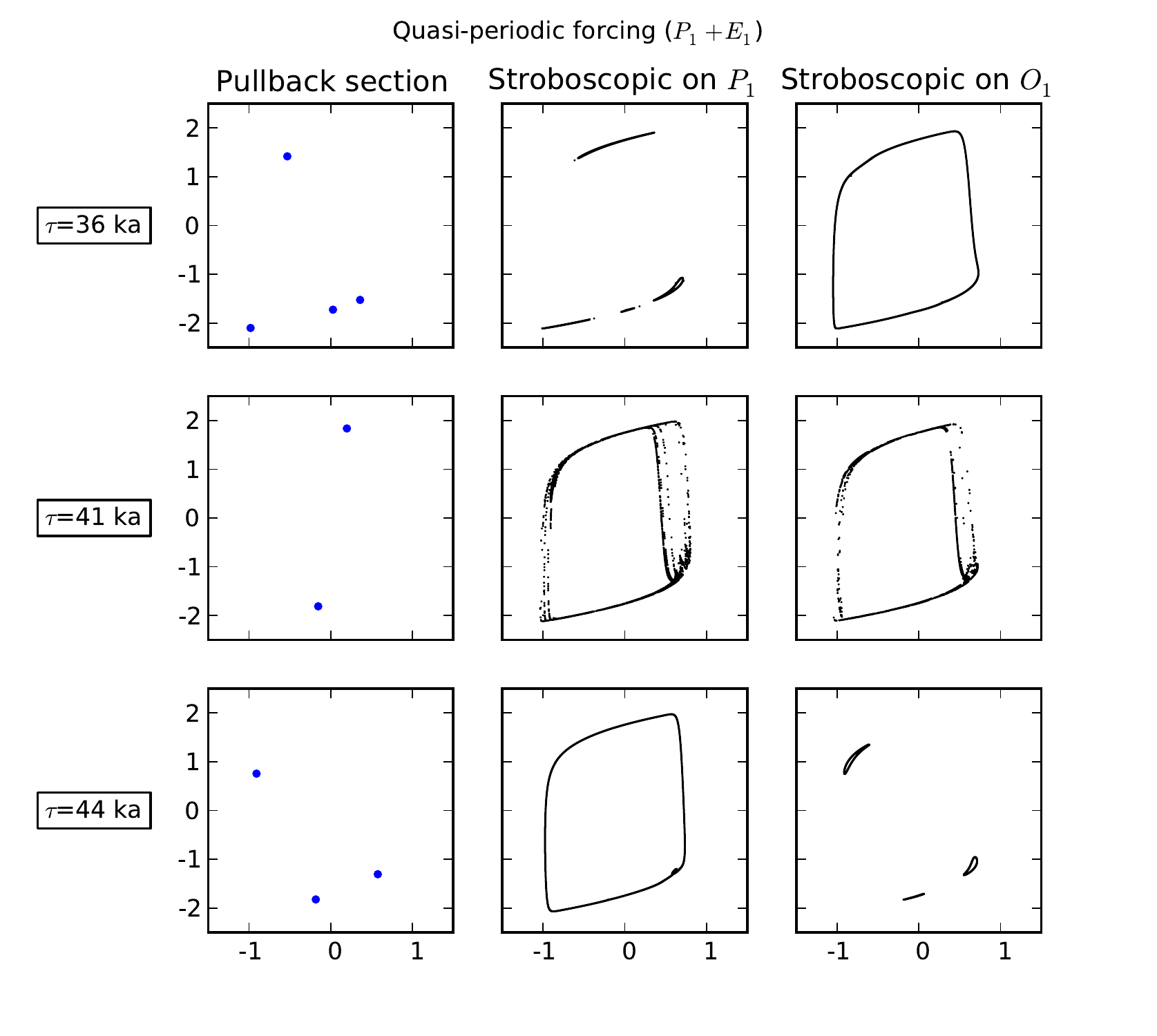}
\caption
{
Pullback (at $t_0=0$) and stroboscopic sections ($t=0+nP_1$ and $t=0+nO_1$) obtained with the van der Pol oscillator with parameters $\alpha=30$, $\beta=0.7$, and forced by $F(t)=\gamma (\sin(2\pi t/P_1+\phi_{P1}) + \sin(2\pi t/O_1)+ \phi_{O1} )$, $P_1=23.7\,\ka$ and $O_1=41.0\,\ka$ and $\gamma=0.6$, and three different values of $\tau$. 
The presence of dots on the pullback section indicates generalised synthronisation. Localised closed curves on the stroboscopic sections indicate frequency locking on the corresponding period (on $P_1$ with $\tau=36\,\ka$  and $O_1$ with $\tau=44\,\ka$), and complex geometries indicate the presence of a strange attractor ($\tau=41\,\ka$). 
}
\label{fig:strob_2periods}
\end{figure}

Let us begin with $\tau=36\, \ka$, which corresponds to a limit cycle in the van der Pol oscillator of period $T_n = 98.79\,\ka$, and consider the stroboscopic section on $P_1$ (\figref{fig:strob_2periods}, line~1). Forcing amplitude $\gamma$ is set to $0.6$. Due to the presence of the $O_1$ forcing, the four points of the periodic case seen on \figref{fig:strob_periodic} have mutated into four local attractors, which appear as closed curves (some are very flat). Every time $P_1$  elapses, the system visits a different local attractor. They are attractors in the sense that they attract solutions of the iteration bringing the system from $t$ to $t+4 \cdot P_1$. 
In this particular example, the system is said to be phase- or frequency-locked on $P_1$ with ratio 1:4 \cite[][p. 68]{Pikovski01aa}, because on average, one ice age cycle takes four precession cycles, even though the response is no longer periodic. 
In this example, the curves on the $P_1$ stroboscopic section nearly touch each other. This implies that synchronisation is not reliable since a solution captured by one of these attractors could easily escape and fall into the basin of attraction of another local attractor. 
One can also see that it is not synchronised on $O_1$ since the stroboscopic section of period $O_1$ shows one closed curve englobing all possible phases. 
It may also be said that the system is synchronised in the \textit{generalised} sense \cite[][p. 150]{Pikovski01aa}, because the pullback section is made of only four points : starting from arbitrary conditions, the system converges to only a small number of  solutions at any time $t$.  It is also stable in the Lyapunov sense, a point that will not be further discussed here, but see \citet{De-Saedeleer12aa}. 

Consider now $\tau=41\, \ka$. The four closed curves on the P1-stroboscopic section have collided and merged into one attractor with strange geometry. 
A similar figure appears on the E1-stroboscopic section. 
The phenomenon of strange non-chaotic attractor has been described since \cite{Romeiras87aa},  its occurrence in the van der Pol oscillator is discussed in \citet{Kapitaniak90ab}, and its relevance to climate dynamics was suggested by \cite{Sonechkin01aa}. 
In our specific example, the system is neither frequency-locked on $P_1$ nor on $O_1$, but it  is synchronised in the generalised sense: the pullback section has two points. Finally, with $\tau = 44\,\ka$ there is frequency-locking on $O_1$ (regime 3:1) but not on $P_1$. 

\begin{figure}[t]
\includegraphics{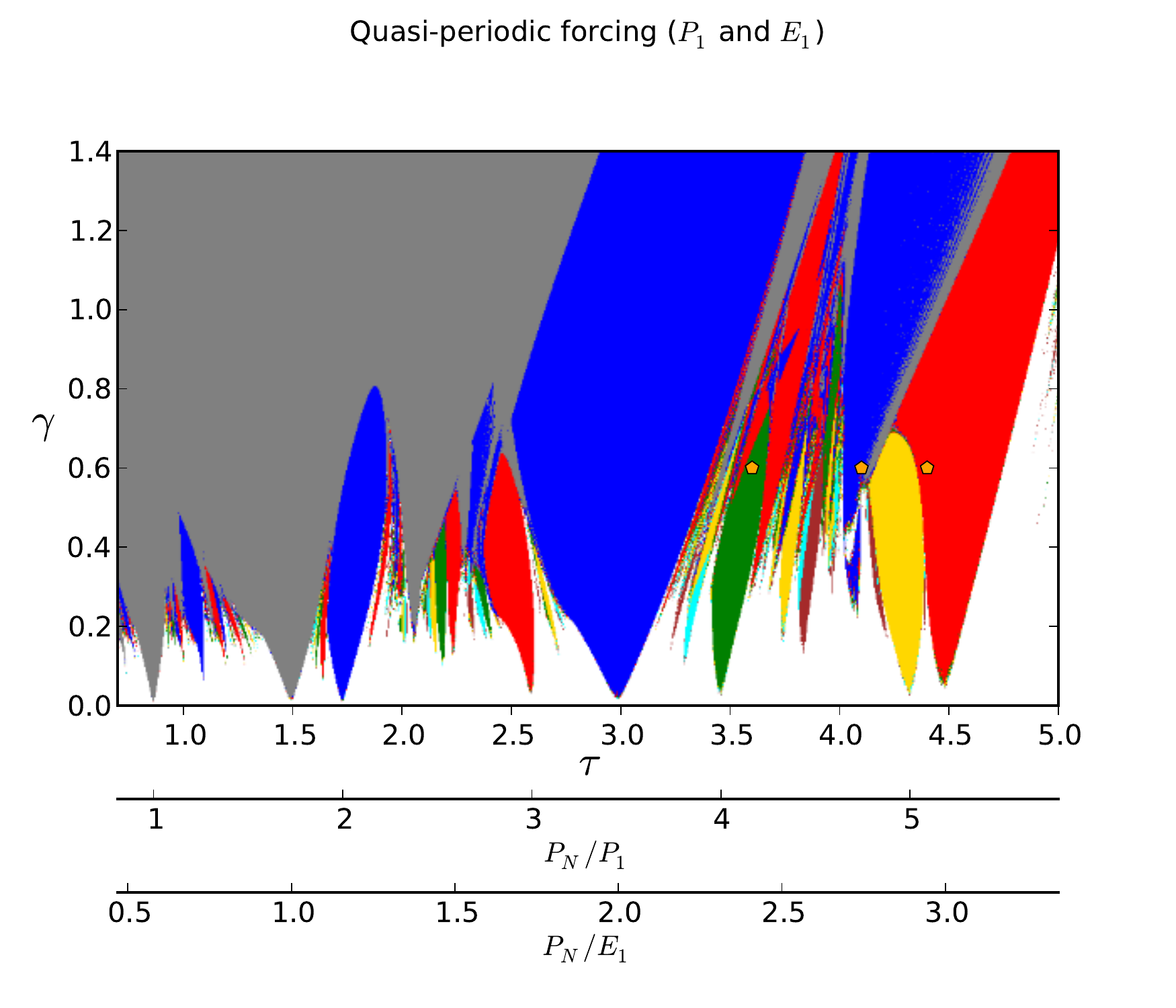}
\caption
{
As Figure \ref{fig:card_periodic} but with a 2-period forcing : 
$F(t)=\gamma ( \sin( 2\pi / P_1 t + \phi_{P1}) +  \sin( 2\pi / O_1 t + \phi_{O2}) )$ (see text for values). Tongues originating from frequency-locking on individal periods merge and give rise to strange non-chaotic attractors. Orange dots correspond to the cases shown on Figure \ref{fig:strob_2periods}. }
\label{fig:card_2periods}
\end{figure}

Clearly, the system  underwent changes in synchronisation regimes as $\tau$ increased from $36$ to $44\, \ka$. 
Further insight may be had by considering the $\tau-\gamma$ plot \figrefp{fig:card_2periods}. The frequency locking regime on $P_1$ lies in the relic of the 1:4 tongue visible in the periodic forcing case \figrefp{fig:card_periodic}. Frequency locking on $O_1$ belongs to the 1:3 tongue associated with $O_1$. The strange non-chaotic regime occurs where the tongues associated with these different forcing components merge.

The word bifurcation has been defined for non-autonomous dynamical systems \cite[chap. 2]{Rasmussen00aa}. 
This is a complex subject and we will admit here the rather informal notion that there is a bifurcation when a local pullback attractor appears or ceases to exist \cite[adapted from Def. 2.42, in ][]{Rasmussen00aa}. With this definition, there is a bifurcation at least every time color changes on \figref{fig:card_2periods} (assuming $t_\mathrm{back}$ is far enough in the past). 

\begin{figure}[t]
\includegraphics{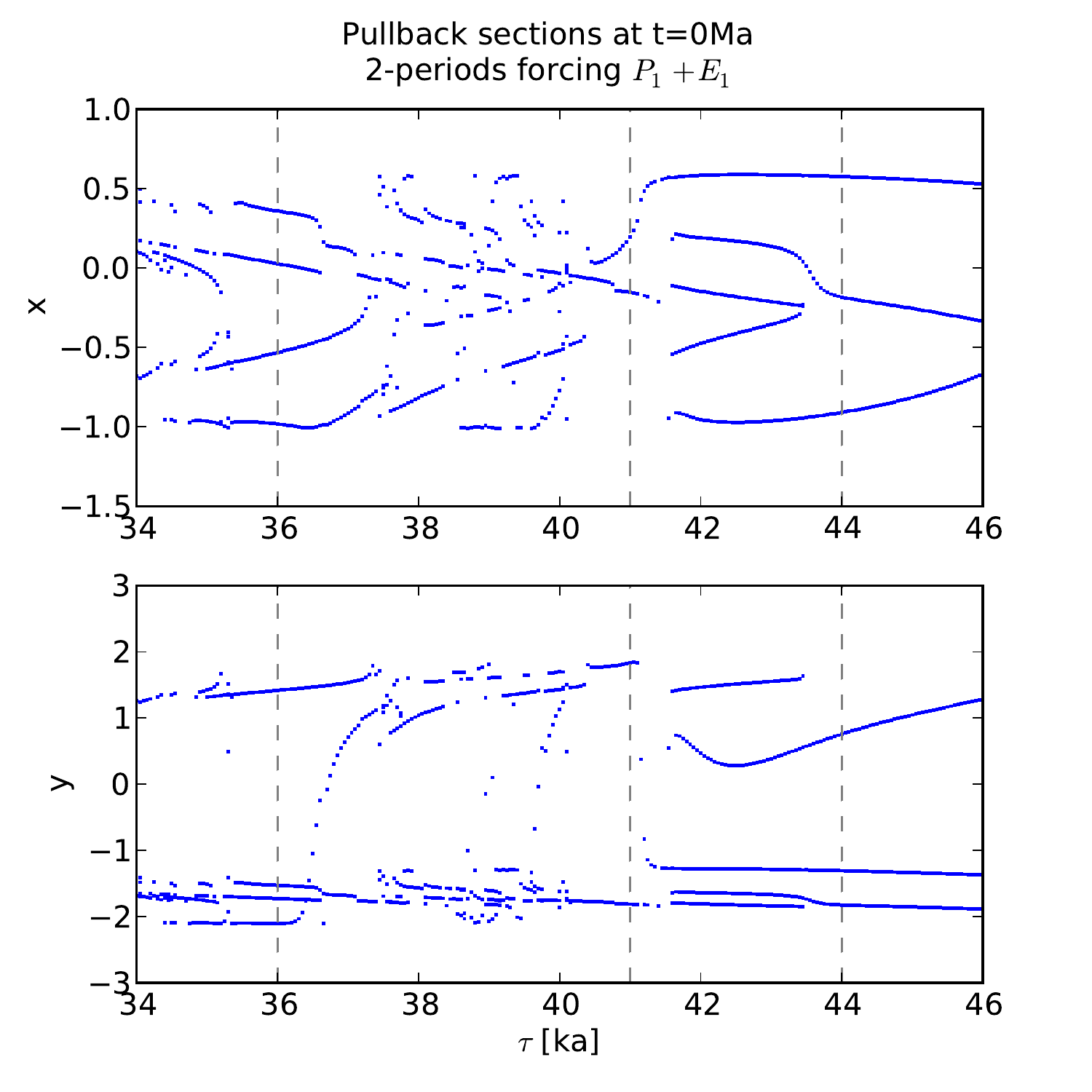}
\caption
{
Pullback solutions of the van der Pol oscillator forced by two periods (as on Figures \ref{fig:strob_2periods} and \ref{fig:card_2periods}) as a function of $\tau$. Forcing amplitude is $\gamma=0.6$ and  the other parameters are as on the previous Figures. Vertical lines indicate the cases shown on Figure \ref{fig:strob_2periods}.
}
\label{fig:bifurc_11}
\end{figure}

Another view on the bifurcation structure may be obtained by plotting the $x$ and $y$ solutions of the system at $t_0=0$, initiated from a grid of initial conditions at $t_{\mathrm{back}}=-5\,\Ma$, as a function of $\tau$, still with $\gamma = 0.6$  \figrefp{fig:bifurc_11}.  This plot outlines a region of weak structural stability, where the system shows pretty strong dependence on the value of the parameters. It  occurs here between $\tau=$ 37 and 42$\,\ka$. 

\label{sect:bifurc11}

These observations have two important consequences for our understanding of the phenomena illustrated on \figref{fig:figure1}.
To see this it is useful to refer to general considerations about autonomous dynamical systems. A bifurcation generally separates two distinct (technically: non-homeomorphic) attractors, which control the asymptotic dynamics of the system. As the bifurcation is being approached, the convergence to the attractor is slower, while the attractor that exists on the other side of the bifurcation may already take some temporary control on the transient dynamics of the system. This is, namely, one possible mechanism of excitable systems. One sometimes refers to `remnant' or `ghost attractors' to refer to these attractors that exist on the other side of the bifurcation and may take control on the dynamics of the system over significant time intervals \cite[e.g.][p. 206]{Nayfeh04aa}

\begin{figure}[t]
\begin{center}
\includegraphics{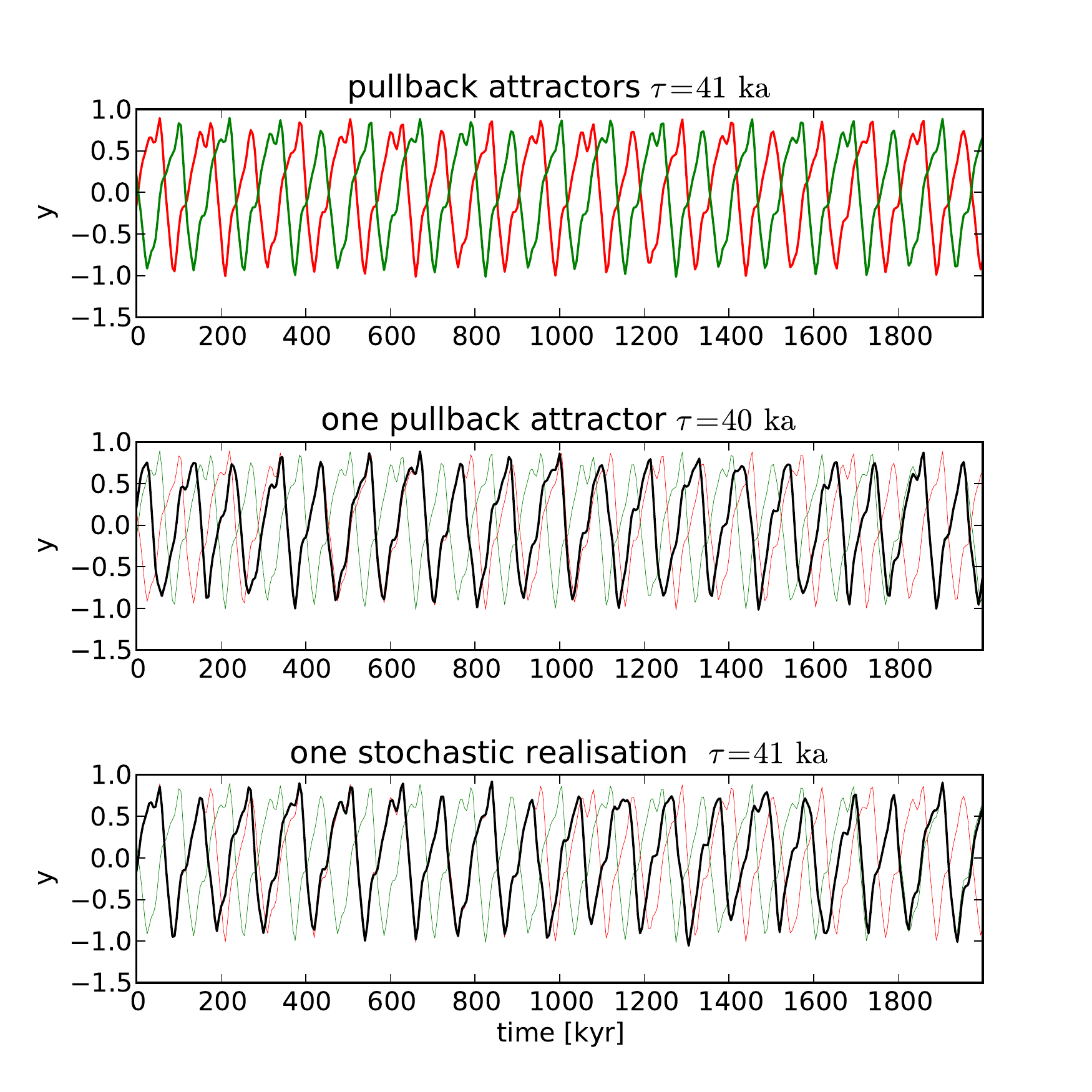}
\end{center}
\caption
{
(top:) Pullback attractors of the forced van der Pol oscillator
 ($\beta=0.7$, $\alpha=30$, $\gamma=0.6$ with 2-period forcing) as on Figure \ref{fig:card_2periods}, 
 for $\tau=41$\,ka. They are reproduced on the graphs below (very thin lines), overlain by (middle:)  
 one pullback attractor with same parameters but $\tau=40\,\mathrm{ka}$, and
 (bottom:) one stochastic realisation of the stochastic van der pol oscillator with $\tau=41\ka$.
}
\label{fig:pullback_vdp}
\end{figure}

The idea may be generalised to non-autonomous systems. 
Consider \figref{fig:pullback_vdp}. The upper plot 
shows the two local pullback attractors of the system obtained with $\tau=41\,\ka$. 
 The middle panel displays one local attractor obtained with $\tau=40$. The  two $\tau=41$ attractors are reproduced with thin lines for comparison. Observe that this $\tau=40$ attractor is qualitatively similar to the $\tau=41$ attractors, and most of its time is spent on a path that is nearly undistinguishable from those obtained with $\tau=41$. However, on a portion of the time interval displayed  it follows a sequence of ice ages that is distinct from those obtained with $\tau=41$. In fact, there are four pullback attractors at $\tau=40$, which clarifies the fact that there is at least one bifurcation between $\tau=40$ and $\tau=41$~ka.

Let us now consider a third scenario. Parameter $\tau=41$, but  an additive stochastic term ($\sigma \ddt{\omega}, \sigma^2=0.25\mathrm{ka}^{-1}$, and $\omega$ symbolises a Wiener process) is added to the second system equation. This is thus a slightly noisy version of the original system. Shown here is one realisation of this stochastic equation, among the infinity of solutions that could be obtained with this system. Expectedly, the  system spends large fractions of time near one or the other of the two pullback attractors. However, it also spends a significant time on a distinct path. Speculatively, this distinct path is under the influence of a `ghost' pullback attractor. As the bifurcation structure is complex and dense, as shown on \figref{fig:bifurc_11}, we expect a host of ghosts to lie around, ready to take control of the system over significant fractions of time, and this is what happens in this particular case.

\begin{figure}[t]
\begin{center}
\includegraphics{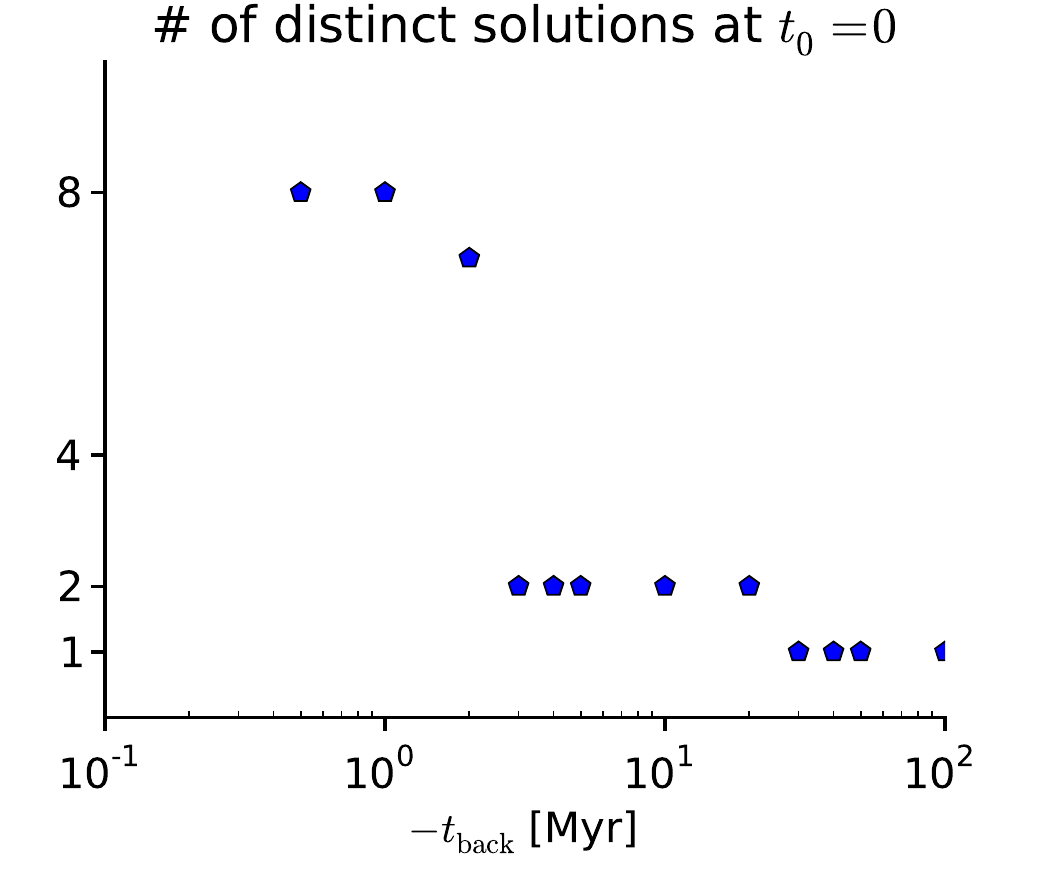}
\end{center}
\caption
{
Number of distinct solutions simulated with the van der Pol oscillator ($\beta=0.7$, $\alpha=30$, $\tau=41$~ka and $\gamma=0.6$ with 2-period forcing as on Figure \ref{fig:card_2periods}, as a function of the time $t_{\mathrm{back}}$ at which 121 distinct initial conditions are considered. The actual stable pullback attracting set(s), in the rigorous mathematical sense, is (are) found for $t_\mathrm{back} \rightarrow -\infty$. 
}
\label{fig:convergence}
\end{figure}

To further support this hypothesis, consider a second experiment. 
\figref{fig:convergence} displays the number of distinct solutions counted at time $t_0=0$, when the system is started from 121 distinct initial conditions at a time back $t_\mathrm{back}$, as a function of $t_\mathrm{back}$. Surprisingly, one needs to go back to $-30$~Ma (million years) to identify the true pullback attractor. Obviously $30$~Ma is a very long time compared to the Pleistocene and so this solution is in practice no more or relevant than the 4 or 8 solutions that can be identified by only starting the system back in time $1$ or $2$~Ma ago. They may be interpreted as ghost (`almost alive') pullback attractors, and following the preceding discussion they are likely to be visited by a system forced by large enough random external fluctuations.

\subsubsection{$P_2 = 22.427\,\ka$ \ and $P_3 = 18.976\,\ka$}
The two periods now being combined are the second and third components of precession, still according to \cite{berger78}. These two periods were selected for two reasons. The first one is that the addition of the two periodic signal produces an interference beating with period $123\, 319\,$yr, not too far away from the usual 100-ka cycle that characterises Late Pleistocene climatic cycles. Second, the period of the beating is not close to an integer number of the two original periods (this occurs, accidentally, when using $P_1$ and $P_3$). This was important to be able to clearly distinguish a synchronisation to the beating from a higher resonance harmonic to either forcing components. 

It is known from astronomical theory that the periodicity of eccentricity  is mechanically related to the beatings of the precession signal \citep{berger78}.
The scientific question considered here is whether the correspondence between the period of ice age cycles and eccentricity is coincidental, or whether a phenomenon of synchronisation of climate on eccentricity developed.

To address this question we need a marker of synchronisation on the precession beating. The Rayleigh number has already been used to this end in palaeoclimate applications \citep{huybers04Pleistocene,  Lisiecki10aa}. Let $P_b$ be the beating period, and $X_i$ the system state snapshotted every $t=t_0 + iP_b$, the Rayleigh number $R$ is defined as $|\sum X_i - \bar X|/\sum| X_i - \bar X|$, where the overbar denotes an average. $R$ is strictly equal to $1$ when the solution is synchronised with a periodic forcing of period $P_b$, assuming no other source of fluctuations. As a reference, \cite{Lisiecki10aa} estimated  0.94 the Rayleigh number of a stacked benthic $\delta^{18}O$ signal with respect to eccentricity over the last million years.

\begin{figure}[t]
\includegraphics{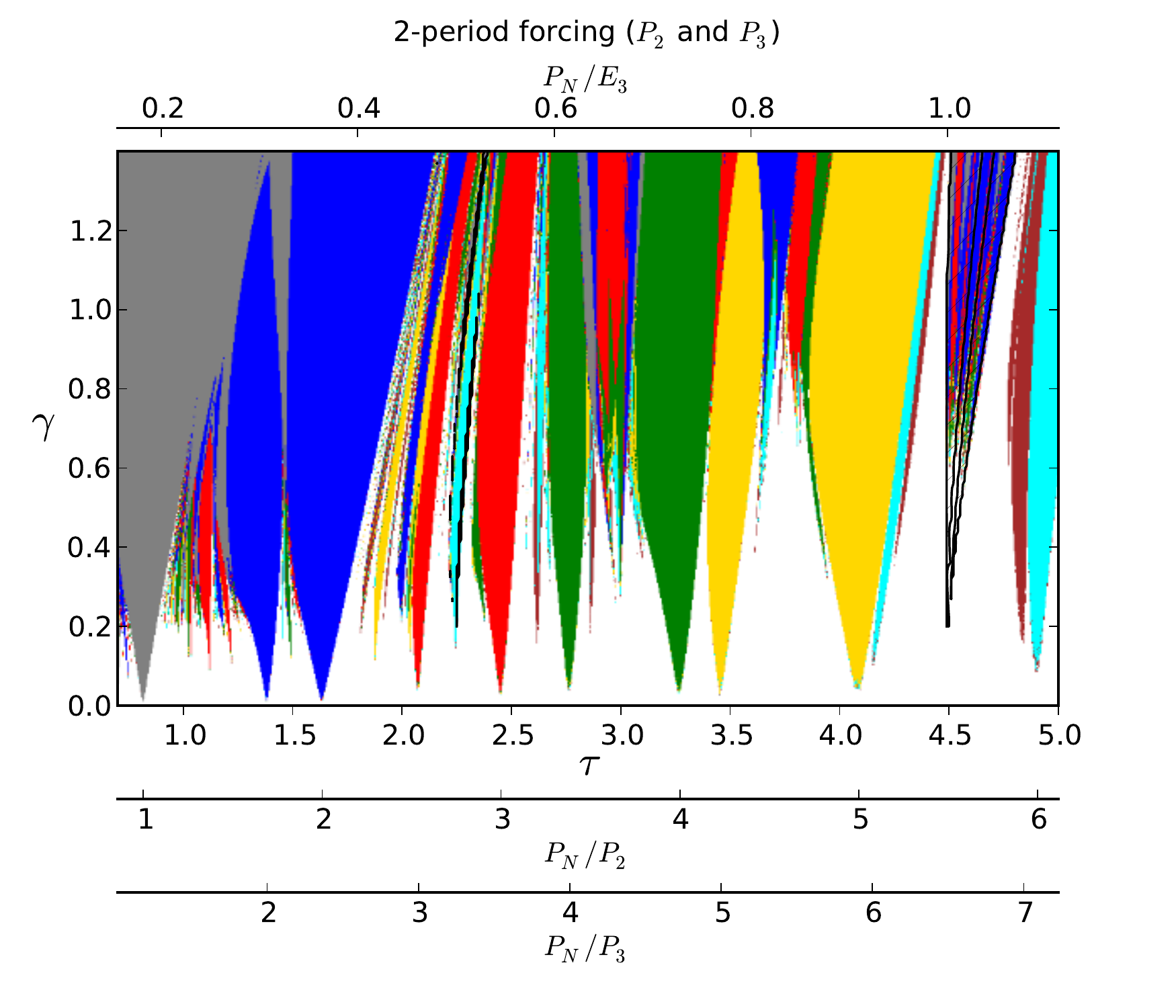}
\vskip-2em
\caption
{
As Figure \ref{fig:card_2periods} but with  : 
$F(t)=\gamma ( \sin( 2\pi / P_2 t + \phi_{P2}) +  \sin( 2\pi / P_3 t + \phi_{P3}) )$. 
Hashes indicate regions of Rayleigh number $>0.95$ on the beating associated with $P_2$ and $P_3$, 
the period of which is called $E_3$ in the Berger (1978) nomenclature (third component of eccentricity).} 
\label{fig:card_precess}
\end{figure}

The bifurcation diagram showing the number of pullback solutions is displayed on \figref{fig:card_precess}. The frequency-locking tongues on $P_2$ and $P_3$ are easily identified at low forcing amplitude; as forcing amplitude increases the tongues merge and produce generalised synchronisation regimes. Regions of synchronisation on $P_b$, identified as $R>0.95$, are hashed. They occur when the system natural period is close to $P_b$ but narrow bands also appear near $T_N=P_b/2$. Observe also that these synchronisation regimes are generally not unique (several pullback attractors co-exist), and  additional sensitivity experiments show that convergence is quite slow. More specifically, the synchronisation diagram was computed here using $t_\mathrm{back}=-10$~Ma. With shorter backward time horizons, the number of remaining solutions identified in the beating-synchronisation regime often exceeds 6 and could not been seen on the graph, while the Rayleigh number was still beyond 0.95. Hence, a high Rayleigh number is not necessarily a good indicator of reliable synchronisation. 

\subsection{Full astronomical forcing \label{sect:fullastr}}
The next step is to consider the full astronomical forcing, as the sum of standardized climatic precession ($\Pi$) and the deviation of obliquity with respect to its standard value ($O$):
\begin{equation}
F(t) = \gamma_p \Pi (t) + \gamma_o O(t),
\label{eq:insol_vdp}
\end{equation}
where
\begin{eqnarray}
\Pi(t) &=& \sum_{i=1}^{N_p} a_i \sin(\omega_{p_i} t + \phi_{p_i})  / a_1
\label{eq:insol_pi}
\\
O &=& \sum_{i=1}^{N_o} b_i \cos(\omega_{o_i} t + \phi_{o_i})  / b_1
\label{eq:insol_eps}
\end{eqnarray}

The various coefficients are taken from \cite{berger78}.  We take $N_p=N_o=34$, so that the signal includes in total 68 harmonic components. With this choice the BER78 solution \citep{berger78} is almost perfectly reproduced. BER78 is still used in many palaeoclimate applications. Compared to  a state-of-the-art solution such as La04 \citep{Laskar04}, the error on amplitude is between 0 and 25 \%, and the error on phase is generally much less than 20$^\circ$.

\begin{figure}[t]
\includegraphics{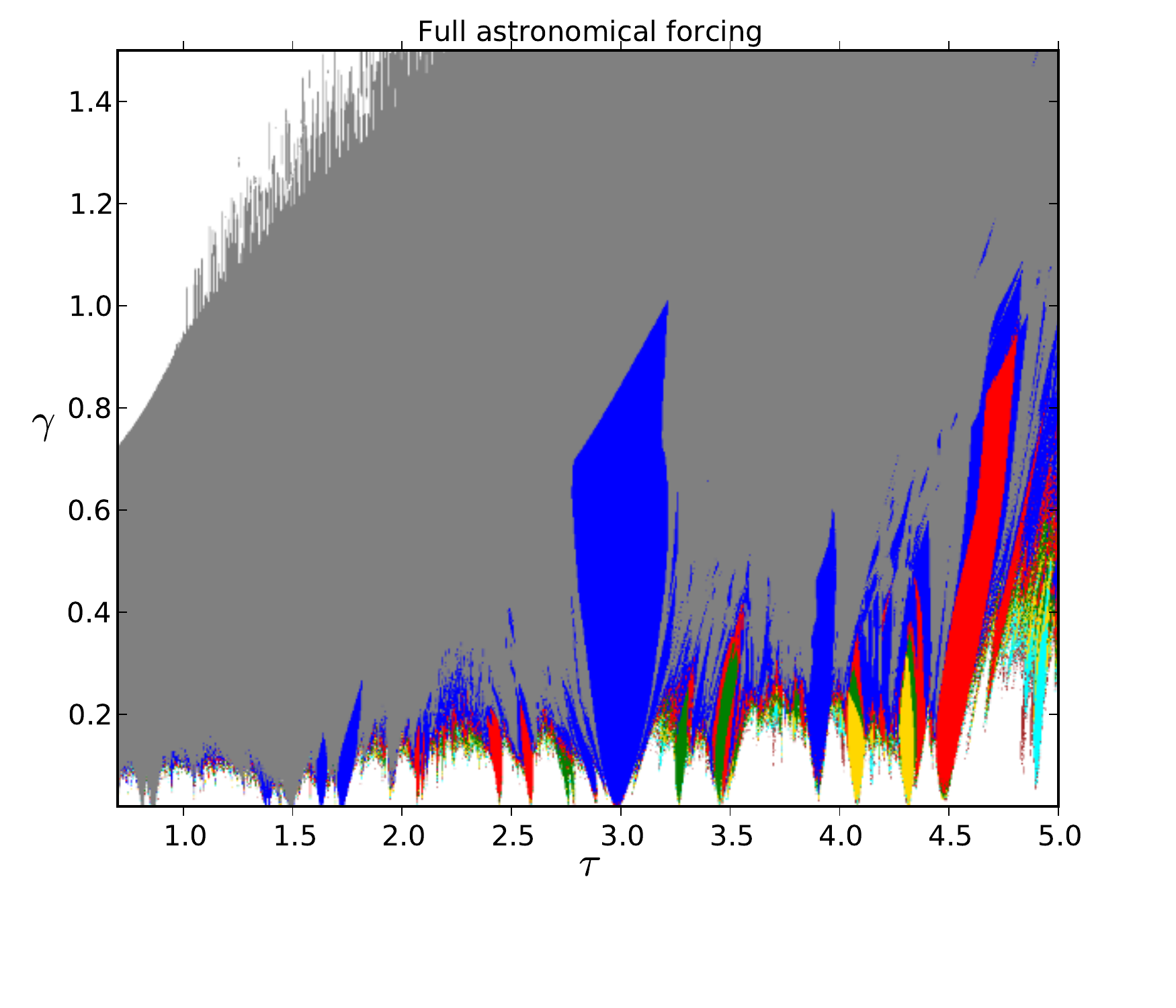}
\caption
{
As Figure \ref{fig:card_2periods} but with the full astronomical forcing,
made of 34 precession and 34 obliquity periods (eq. \eqref{eq:insol_vdp}).
The white area in the top-left is a zone of numerical instability caused by inappropriate
time-step.
}
\label{fig:card_full_vdp}
\end{figure}

The bifurcation diagram representing the number of pullback solutions as a function of forcing amplitude and $\tau$ is shown on \figref{fig:card_full_vdp}. We have taken $\gamma = \gamma_p = \gamma_o$. One recognises the tongues originating from the individual components merging gradually into a complex pattern. The number of attractors settles to 1 as the amplitude of the forcing is further increased. Let us call this the 1-pullback attractor regime. We already know that synchronisation is generally not reliable in the region characterised by  the complex and dense bifurcation region, where more than one attractor exist.  The remaining problem is to characterise the reliability of synchronisation in the 1-pullback attractor regime.

\begin{figure}[t]
\includegraphics{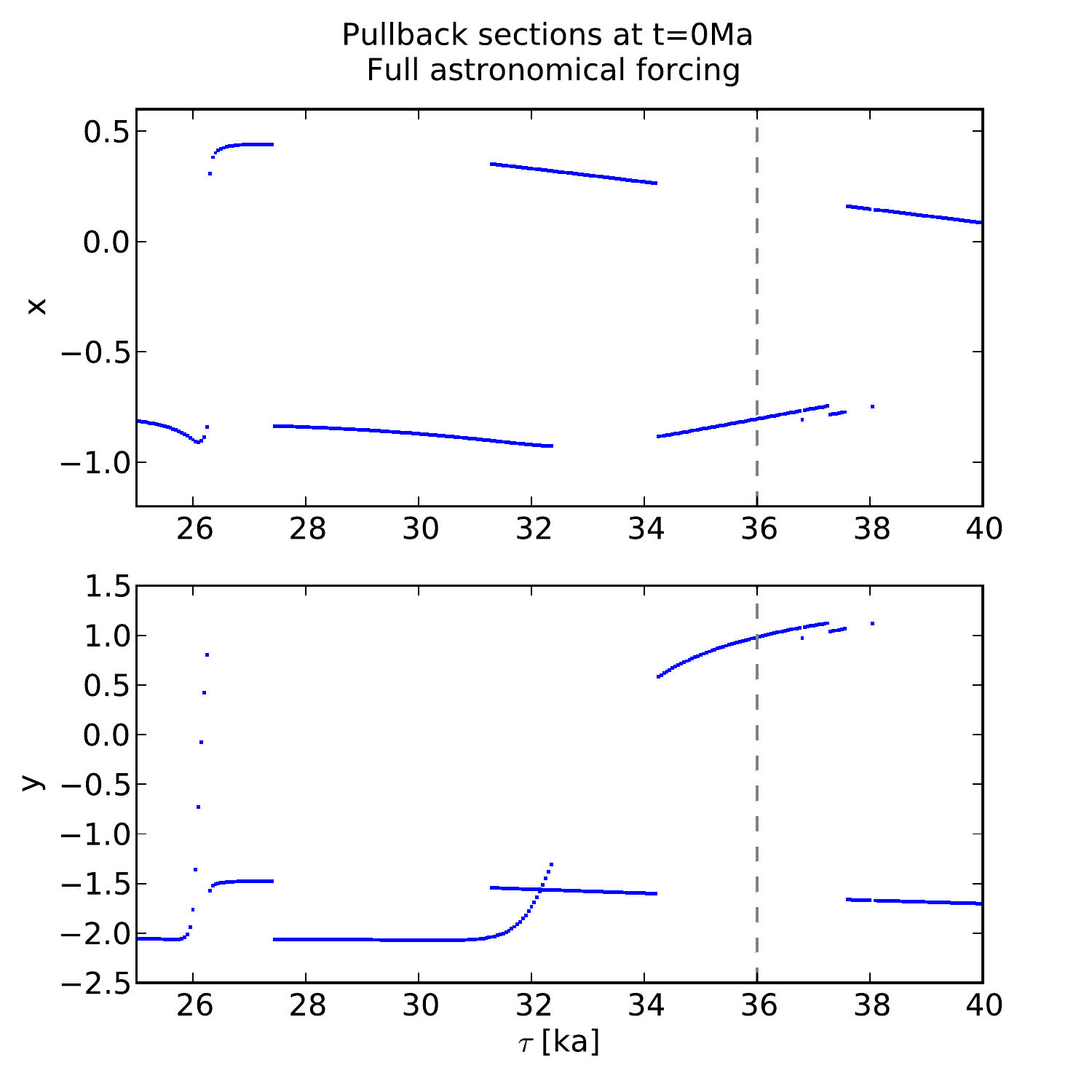}
\caption
{
As Figure \ref{fig:bifurc_11} but with the full astronomical forcing, with parameters as
on Figure \ref{fig:card_full_vdp} and $\gamma=1.0$. 
}
\label{fig:bifurc_34}
\end{figure}

The literature says little about systematic approaches to quantify the reliability of generalised synchronisation with quasi-periodic forcings. To develop further the ideas developed in section \ref{sect:bifurc11}, one can
plot pullback solutions at a certain time $t$ as a function of one or several parameters. This is done \figref{fig:bifurc_34}. 
Here $\gamma$ is kept constant ($=1.0$) and $\tau$ is varied between 25 and 40. There is a brief episode of 2-solution regime between 31 and 32\,ka.  Within the 1-solution regime there is a number of abrupt transitions (at 26, 27, 34 and 37 \,ka).

As we have seen above, the density of bifurcations in the parameter space is an indicator of the structural stability of the system. Changes in the number of pullback attractors are clearly bifurcations. 
Abrupt variations such as near $\tau=26,\ 27, \ 34$ and $37$~ka could well be bifurcations because one observes slower convergence near the transitions, with co-existence of two or more solutions when $t_\mathrm{back}$ is only 1~Ma (not shown).  It may also be verified that near these transition zones scenarios similar to those depicted on \figref{fig:pullback_vdp} may be observed. Most of the transition zones disappear as $\gamma$ is further increased.  Hence, being in the 1-pullback attractor zone is not quite enough to guarantee a reliable synchronisation: one need to be deep into that zone. 

\section{Other models \label{sect:others}}
We now consider  6 previously published models. Mathematical details are given in the Appendix and the codes are available on-line at  
\texttt{https://github.com/mcrucifix}.
\begin{description}
\item[SM90:]  This is a model with three ordinary differential equations representing the dynamics of ice volume, carbon dioxide concentration and deep-ocean temperature. The astronomical forcing is linearly introduced in the ice volume equation, under the form of  insolation at 65$^\circ$ North on the day of summer solstice. Only the carbon dioxide equation is non-linear, and this non-linearity induces the existence of a limit-cycle solution---spontaneous glaciation and deglaciation---in the corresponding autonomous system. The SM90 model is thus a mathematical transcription of the hypothesis according to which the origin 100,000 year cycle is to be found in the biological components of Earth's climate.
\item[SM91:] This model is identical to SM90 except for a difference in the carbon cycle equation. 
\item[PP04:] The Paillard-Parrenin model \citep{paillard04eps} is also a 3-differential-equation system, featuring Northern Hemisphere ice volume, Antarctic ice area and carbon dioxide concentration. The carbon dioxide equation includes one non-linear term associated to a switch on/off of the southern ocean ventilation. Astronomical forcing is injected linearly at three places in the model: in the ice-volume equation, in the carbon dioxide equation, and in the ocean ventilation parameterisation. The autonomous version of the model also features a limit cycle. As in SM90 and SM91 the non-linearity introduced in the carbon cycle equation plays a key role but the bifurcation structure of this model differs from SM90 and SM91 \citep{Crucifix12aa}.
\item[T06:] The \citet{tziperman06pacing} model is a mathematical idealisation of more complex versions previously published by \citet{Gildor-Tziperman-2000:sea}. T06 features the concept of sea-ice switch, according to which sea-ice growth in the Northern Hemisphere inhibits accumulation of snow over the ice sheets, and vice-versa. Mathematically, T06 is presented as a hybrid model, which is the combination of a differential equation in which the astronomical forcing is introduced linearly as a summer insolation forcing term, and a discrete variable, which may be 0 or 1 to represent the absence or presence of sea-ice in the northern hemisphere. 
\item[I11:] The \citet{Imbrie11aa} was introduced by its authors as a ``phase-space" model. It is a 2-D model, of which the equations were designed to distinguish an `ice accumulation phase' and an `abrupt deglaciation' phase, which is triggered when a threshold defined in the phase space is crossed. I11 was specifically tuned to reproduce the phase-space characteristics of the benthic oxygen isotopic dynamics. A particularity of this model is that the phasing and amplitude of the forcing depend on the level of glaciation.
\item[PP12:] Similar to \citet{Imbrie11aa}, the \cite{Parrenin12ab} model distinguishes accumulation and deglaciation phases. Accumulation is a linear accumulation of insolation, without restoring force (hence similar to equation 1 of the van der Pol oscillator); deglaciation accumulates insolation forcing but a negative relaxation towards deglaciation is added. Contrarily to \cite{Imbrie11aa}, the trigger function, which determines the regime change, is mainly a function of astronomical parameters. An ice volume term only appears in the function controlling the shift from `accumulation' to `deglaciation' regime. 
\end{description}

\begin{figure}[t]
\begin{center}
\includegraphics{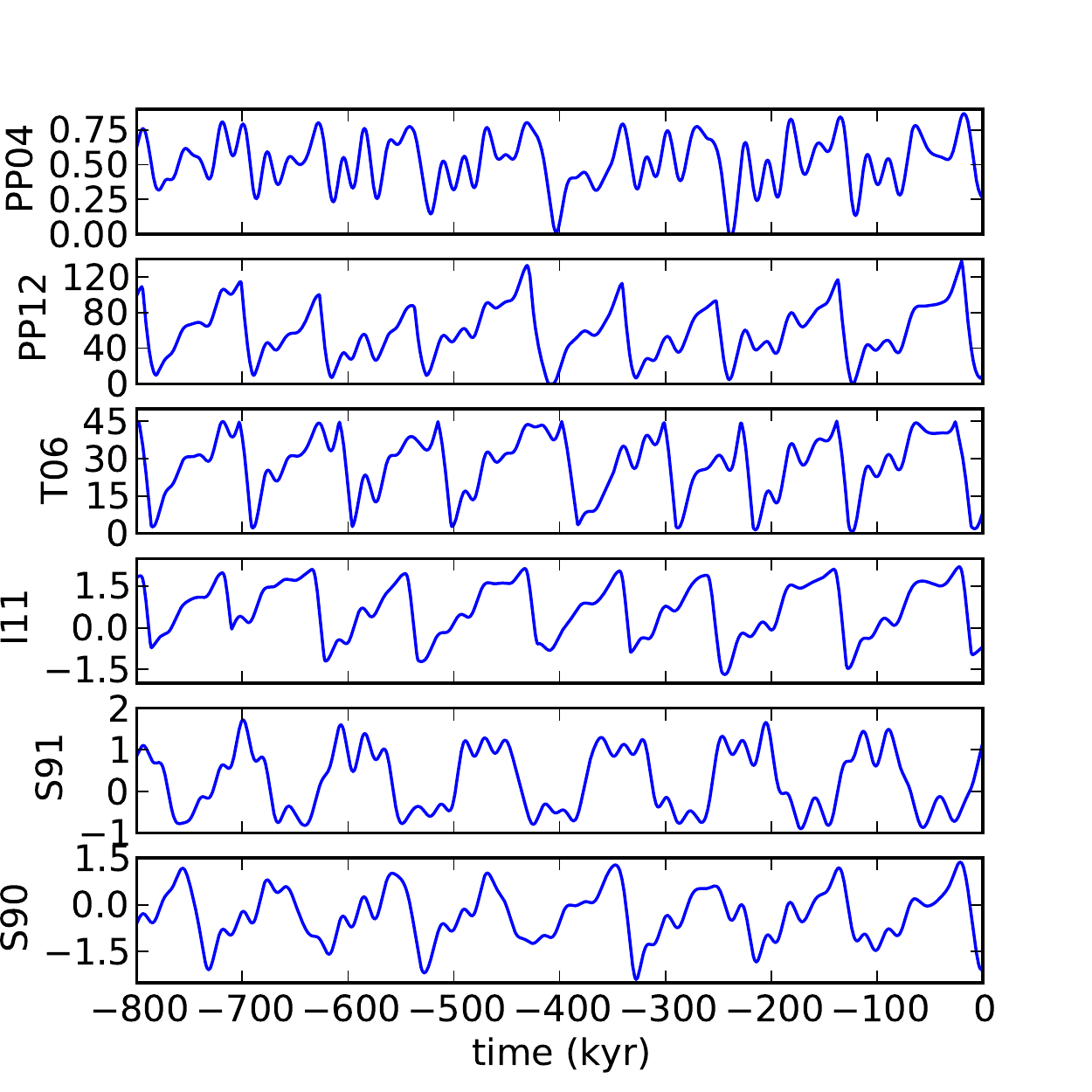}
\end{center}
\caption
{
Pullback attractors obtained for 6 models over the last 800~ka, forced by astronomical forcing with the parameters of the original publications. Shown is the model component representing ice volume. Units are arbitrary in all models, except in T06 (ice volume in $10^{15}\mathrm{m}^{3}$) and PP12 (sea-level equivalent, in $\mathrm{m}$). 
}
\label{fig:pullback_others}
\end{figure}

The ice volumes (or, equivalently, glaciation index or sea-level) simulated by each of these models are shown on \figref{fig:pullback_others}. Shown here are estimates of the pullback attractors. More specifically, the trajectories obtained with an ensemble of initial conditions at $t_{\mathrm{back}}=-20\,$Ma; in some cases the curves actually published (in particular in SM90) are not pullback attractors, but \textit{ghost} trajectories in the sense illustrated on \figref{fig:convergence}. In some cases (PP04 and PP12) the parameters had to be slightly adjusted to reproduce the published version satsifactorily. Details are given in appendix.

Some of these models include as much as 14 adjustable parameters (e.g.: PP04) and a full dynamical investigation of each of them is beyond the scope of the study. Rather, we proceeded as follows.  Every model responds to a state equation, which may be written, in general (assuming a numerical implementation), as:
\begin{eqnarray*}
\left\{
\begin{array}{rcl}
t_{i+1}&=& t_i + \delta t \\
x_{i+1} &=& x_i +  \delta t \ f(x_i, F(t) ),
\end{array}
\right.
\end{eqnarray*}
where $t_i$ is the discretized time, $x_i$ the climate state (a 2-D or 3-D vector) at $t_i$ and $F(t)$ is the astronomical forcing, which is specific to each model because the different authors made different choices about the respective weights and phases of precession and obliquity. 

In all generality, the equation (or its numerical approximation) may be rewritten as follows, posing $\tau=1$ and $\gamma=1$:
\begin{eqnarray*}
\left\{
\begin{array}{rcl}
t_{i+1}&=& t_i + \delta t \\  x_{i+1} &=& x_i + \tau \delta t \ f(x_i, \gamma F(t) ),
\end{array}
\right.
\end{eqnarray*}

The parameters $\tau$ and $\gamma$ introduced this way have a similar meaning as in the van der Pol oscillator, since $\tau$ controls the characteristic response time of the model, while $\gamma$ controls the forcing amplitude. 

\begin{figure}[t]
\includegraphics[scale=0.68]{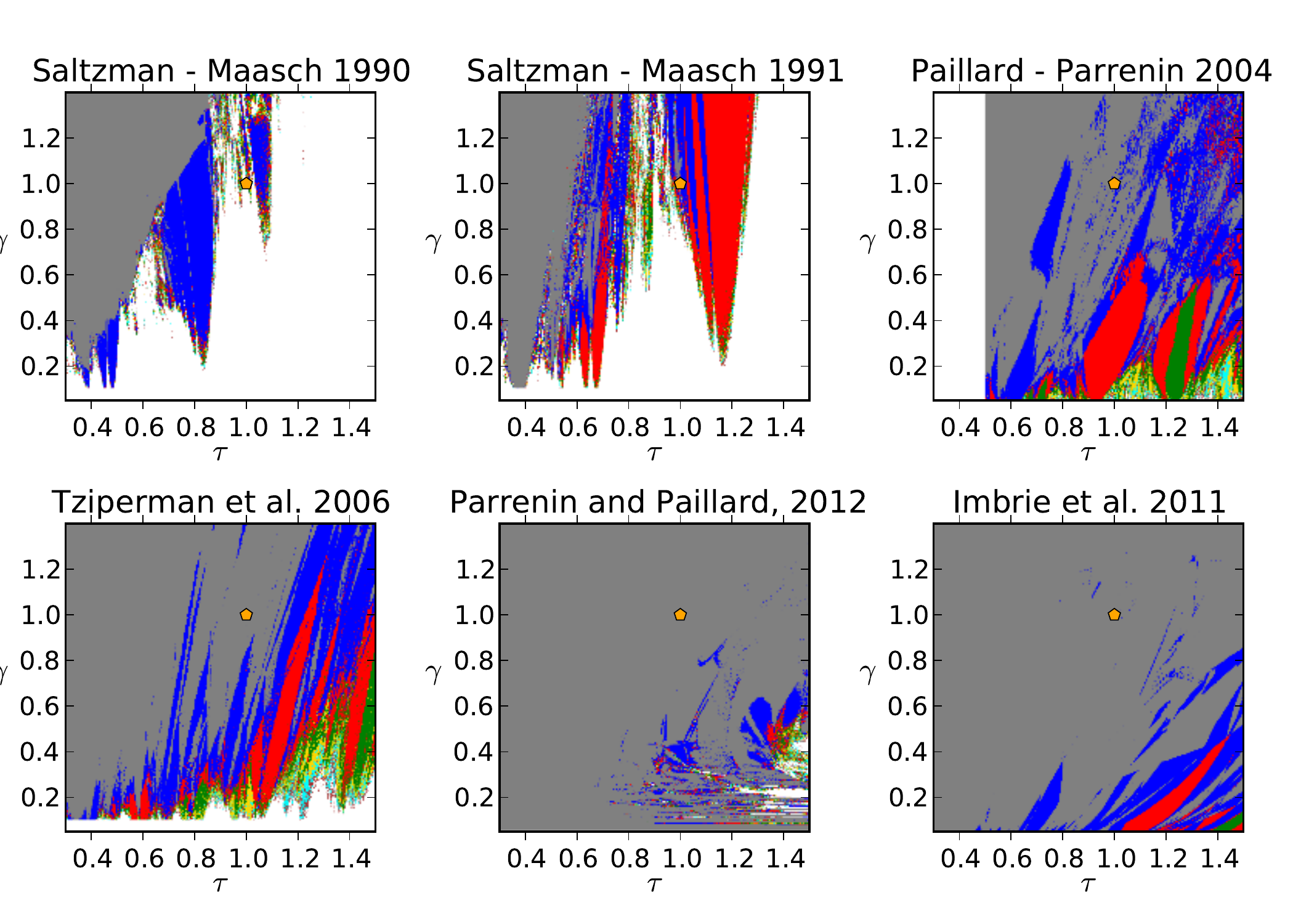}
\caption
{
As Figure \ref{fig:card_full_vdp} but for 6 models previously published. 
Orange dots correspond to standard (published) parameters of these models. 
}
\label{fig:card_others}
\end{figure}

Bifurcation diagrams, similar to \figref{fig:card_full_vdp} are then shown on \figref{fig:card_others}. Remember that $\gamma=\tau=1$ corresponds to the model as originally published.

A first group of four models appears (SM90, SM91, T06 and PP04), on which one recognises a similar tongue-synchronisation structure as in the van der Pol oscillator. This was expected since these models are also oscillators with additive astronomical forcing. Depending on parameter choices synchronisation may be reliable or not. Synchronisation is clearly not reliable in the standard parameters used for SM90 and SM91. In T06 the standard parameters are not far away from the complex multi-pullback-attractor regime and this explains why transitions such as displayed on \figref{fig:figure1} could be obtained under small parameter changes (or, equivalently, with some noise, as discussed in the original \cite{tziperman06pacing} study).  The standard parameter choice of PP04 is further into the stability zone and, indeed, experimenting with this model shows that instabilities such as displayed on \figref{fig:figure1} are harder to obtain. 

\begin{figure}[t]
\includegraphics{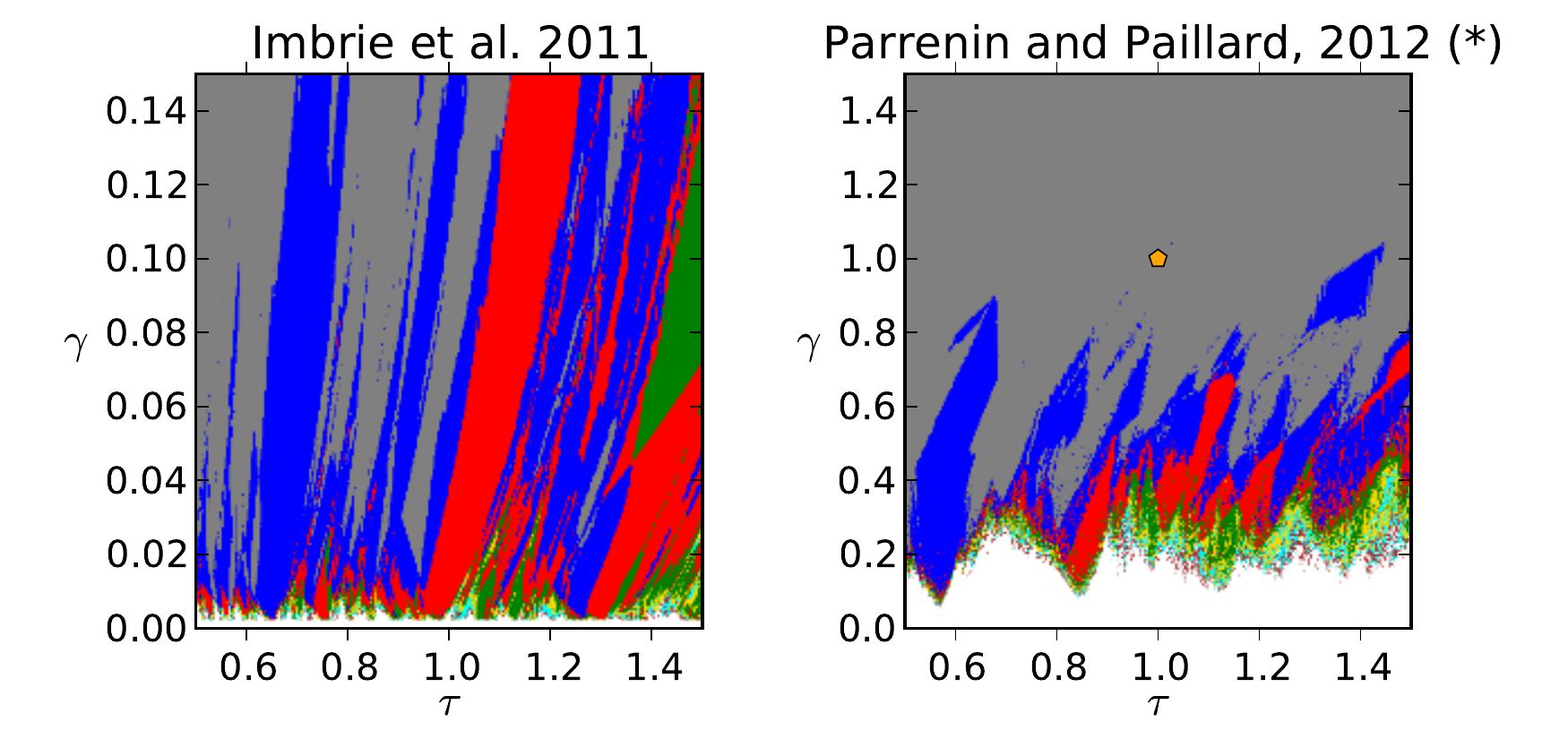}
\caption
{
As Figure \ref{fig:card_others} for (left) the I11 model, but
with a zoom on the $x-$axis , and (right) for a slightly
modified version of the PP12 model, in which the transition
between deglacial and glacial states is also controlled by 
ice volume, as opposed to the original PP12 model. 
}
\label{fig:card_others_sup}
\end{figure}

I11, with standard parameters, is also fairly deep in the stable zone. One has to consider much smaller forcing values than published to recognize the synchronisation tongue structure that characterises oscillators \figrefp{fig:card_others_sup}.

PP12 finally turns out to be the only case not showing a tongue-like structure. This may be surprising because this model also 
has limit cycle dynamics (self-sustained oscillations in absence of astronomical forcing). 
Some aspects of its design resemble the van der Pol oscillator. The role of the variable $y$ in the van der Pol oscillator is here played by the \textit{mode}, which may either be $g$ (glaciation) or $d$ (deglaciation). Also similar to the van der Pol, the direct effect of the astronomical forcing on the ice volume ($x$ in the van der Pol; $v$ in PP04) is additive. T06 has also similar characteristics. 
The distinctive feature of PP12 is that the transition between deglaciation and glaciation modes is controlled by the astronomical forcing and not by the system state, as in the other models. To assess the importance of this element of design we considered a hacked version of PP12, where the $d\rightarrow g$ occurs when for $v<v_1$ (cf. appendix \ref{ssect:pp12} for model details). In this case the tongue synchronisation structure is recovered, with standard parameters marginally in the reliable synchronisation regime
\figrefp{fig:card_others_sup}.

\conclusions

The present article is built around the paradigm of the `pacemaker', that is, the timing of ice ages arises as a combination of climate's internal dynamics with the variations of incoming solar radiation induced by the variations of our planet's orbit and obliquity. This is not the only explanation of ice ages, but this is certainly one of the most plausible. 

In this study we paid attention to the dynamical aspects that may affect the stability of the ice age sequence and its predictability. First, the astronomical forcing has a rich harmonic structure. We showed that a system like the van der Pol oscillator is 
more likely to be synchronised on the astronomical forcing as Nature provides it than on a periodic forcing, because the fraction of the parameter space corresponding to synchronisation is larger in the former case. A synchronised system is Lyapunov stable, so that at face value this would imply that the sequence of ice ages is stable. However, ---this is the second point--  even if the dynamical structure of the Pleistocene climate was correctly identified, there would be at least two sources of uncertainties~: random fluctuations associated with the chaotic atmosphere and ocean and other statistically random forcings like volcanoes; and  uncertainty on system parameters. In theory both types of uncertainties point to different mathematical concepts: path-wise stability to random fluctuations in the first case, and the structural stability in the second. In practice, however, lack of either form of stability will result in similar consequences: quantum skips of insolation cycles in the succession of ice ages. This was the lesson of \figref{fig:pullback_vdp}. 

It was shown here that, compared to periodic forcing, the richness of the harmonic structure of astronomical forcing favours situations of weak structural stability. To preserve stability, the richness of the astronomical forcing has to be compensated for by large enough forcing amplitude. 

Out of the seven models tested here, we ignore which one best captures ice ages dynamics. The overwhelming complexity of the climate system does not allow us to securely select the most plausible model on the sole basis of our knowledge of physics, biology and chemistry. 
Consequently, while we have understood here how and why the sequence of ice ages could be unstable in spite of available evidence (astronomical spectral signature; Rayleigh number), estimating the stability of the sequence ice ages and quantifying our ability to predict ice ages is also a problem of statistical inference : calibrating and selecting stochastic dynamical systems based on both theory and observations, which are sparse and characterised by chronological uncertainties. 
A conclusive demonstration of our ability to reach this objective is still awaited.

\appendix

\section{Insolation \label{sect:appinsol}}
In the following models, the forcing is computed as a sum of precession ($\Pi := e\sin\varpi/a_1$), co-precession ($\tilde \Pi := e\cos\varpi/a_1$) and obliquity ($O := (\varepsilon - \varepsilon_0)/b_1$) computed according to the \cite{berger78} decomposition (Figure \ref{fig:astro}, and eqs. (\ref{eq:insol_pi} -- \ref{eq:insol_eps})). More precisely, we use here these quantities scaled ($\bar P$, $\bar{\tilde\Pi}$ and $\bar O$) such as they have unit variance. 
All insolation quantities used in climate models may be approximated as a linear combination of $\bar\Pi$, $\bar{\tilde\Pi}$ and $\bar O$. For example:
\begin{itemize}
\item Normalised summer solstice insolation at 65$^\circ$N = $0.8949 \bar \Pi + 0.4346 \bar O$
\item Normalised  insolation at 60$^\circ$ S on the 21st February  = $-0.4942 \bar \Pi + 0.8399  \bar{\tilde\Pi} + 0.2262 \bar O$
\end{itemize}
\section{Model definitions \label{sect:app_models}}
\subsection{SM90 model}
\begin{eqnarray*}
\ddt{x} &=& -x - y - v\, z - u F(t)\\
\ddt{y} &=& -p z + r \, y + s \, z^3 - w\, y \, z - z^2\, y \\
\ddt{y} &=& -q \, (x + z)
\end{eqnarray*}
$p=1.0$, $q=2.5$, $r=0.9$, $s=1.0$, $u=0.6$, $v=0.2$ and $w=0.5$,  and $F(t)$ is (here) insolation on the day of summer solstice, at 65$^\circ$ North, normalised  (results are qualitatively insensitive to the exact choice of insolation). 
\subsection{SM91 model}
\begin{eqnarray*}
\ddt{x} &=& -x - y - v\, z - u F(t)\\
\ddt{y} &=& -p z + r \, y - s \, y^2  -  y^3 \\
\ddt{y} &=& -q (x + z)
\end{eqnarray*}
$p=1.0$, $q=2.5$, $r=1.3$, $s=0.6$, $u=0.6$ and $v=0.2$,  and $F(t)$ is (here)  insolation on the day of summer solstice, at 65$^\circ$ North, normalised  (results are qualitatively insensitive to the exact choice of insolation). 
One time unit is 10~ka. 
\subsection{PP04 model}
The three model variables are $V$ (Ice volume), $A$ (Antarctic Ice Area) and $C$ (Carbon dioxide concentration):
\begin{eqnarray*}
\ddt{V}&=&  \frac{1}{\tau_V} ( - x C -y F_1(t) + z - V ) \\
\ddt{A}&=&  \frac{1}{\tau_A} ( V - A ) \\
\ddt{C}&=&  \frac{1}{\tau_C} ( \alpha F_1(t) - \beta V + \gamma H + \delta - C),
\end{eqnarray*}
$\tau_V=15\,\ka$,  $\tau_C=0.5\,\ka$, $\tau_A=1.2\,\ka$, $x=1.3$, $y=0.4$ (was 0.5 in the original paper),  $z=0.8$, $\alpha=0.15$, $\beta=0.5$, $\gamma=0.5$,  $\delta=0.4$,  $a=0.4$, $b=0.7$, $c=0.01$,  $d=0.27$; $H=1$ if $aV-bA+d-c\,F_2(t) < 0$, and $H=0$ otherwise. $F_1(t)$ is the normalised, summer-solstice insolation at 65$^\circ$ North, and $F_2(t)$ is insolation at 60$^\circ$ S on the 21st February (taken as 330$^\circ$ of true solar longitude). Other quantities ($V$, $A$, $C$) have arbitrary units. 
\subsection{T06 model}
The two model variables are $x$ (ice volume) and $y$ (sea-ice area).  $x$ is expressed in units of $10^{15}\, \mathrm{m}^3$. 
\providecommand{\asi}{\alpha_{\mathrm{si}}}
\begin{eqnarray*}
\ddt{x}&=& ( p_0 - K\,x ) (1 - \asi) - (s + s_m F(t))
\end{eqnarray*}
The equation represents the net ice balance, as accumulation minus ablation,  and $\asi$ is the sea-ice albedo. $\asi=0.46 y$. $y$ switches from 0 to 1 when $x$ exceeds $45\times 10^6\,\mathrm{km}^3$, and switches from 1 to 0 when $x$ decreases below $3\times 10^6\,\mathrm{km}^3$. The parameters given by \citet{tziperman06pacing} are: $p_0=0.23$Sv, $K=0.7/(40\,\ka)$, $s=0.23$Sv and $s_m=0.03$Sv, where 1 Sv=$10^6$m$^3$/s. $F(t)$ is the normalised, summer-solstice insolation at 65$^\circ$ North.
\subsection{I11 model}
Define first:
\begin{eqnarray*}
\phi &=& \frac{\pi}{180} \cdot  
\left\{
\begin{array}{ll}
10- 25y \quad & \textrm{where } y<0,  \\
10 \quad  & \textrm{elsewhere.}
\end{array}
\right. \\
\theta &=& \left\{
\begin{array}{ll}
0.135 + 0.07  y \quad & \textrm{where } y<0,  \\
0.135 \quad \textrm{elsewhere.}
\end{array}
\right. \\
a &=& 0.07 + 0.015 y \\
h_O  &=& (0.05 - 0.005 y ) \bar O \\
h_\Pi  &=& a \bar\Pi \cdot \sin\phi \\
h_{\tilde\Pi}  &=& a \bar{\tilde\Pi} \cos\phi  \\
F &=& h_\Pi  + h_{\tilde\Pi}  + h_O \\
d &=&  -1 + 3 \frac{(F + 0.28)}{ 0.51 } \\
\end{eqnarray*}
With these definitions:
\begin{equation}
\left\{
\begin{array}{ll}
\ddot{y} &=  0.5  [0.0625  (d - y) - \frac{\dot{y}^2}{d-y} ]  \quad \textrm{where } \dot{y} > \theta, \\
\dot{y} &= -0.036375 + y  [ 0.014416 + y    (0.001121 - 0.008264y)] + 0.5F \quad \mathrm{elsewhere,} \\
\end{array}
\right. 
\end{equation}
where $\dot{y} := \ddt{y}$ and $\ddot{y} := \ddt{\dot{y}}$. 
Note that the polynomial on the right-hand-side of the equation for $\dot y$ is a continuous fit to the piece-wise function used in the original \cite{Imbrie11aa} publication.
Time units are here $\ka$. 
\subsection{PP12 model \label{ssect:pp12}}
This is a hybrid dynamical system, with ice volume $v$ (expressed in $m$ of equivalent sea-level) and state, which may be $g$ (glaciation) or $d$ (deglaciation). 

Define first 
\begin{equation}
f(x):=
\left\{
\begin{array}{l}
x + \sqrt{4 a^2 + x^2} - 2a \quad \textrm{where } x>0, \\
x  \quad \textrm{elsewhere }. \\
\end{array}
\right.
\end{equation}
with $a=0.68034$.
Then define the following quantities, standardized as follows:
\begin{eqnarray*}
\Pi^\star &=&  (f(\bar\Pi)  - 0.148 ) / 0.808 \\
\tilde \Pi^\star &=&  (f(\bar{\tilde \Pi})  - 0.148 ) / 0.808 
\end{eqnarray*}
the threshold $\theta= k_\Pi \bar\Pi + k_{\tilde\Pi} \tilde{\bar\Pi} + k_O  \bar{O}  $,
and finally the following rule controlling the transition between state $g$ and $d$:
\newline
\begin{displaymath}
\left\{  \parbox{\fill}{ \vskip-1em
\begin{eqnarray*}
d \rightarrow g &\quad& \mathrm{if}\quad \theta < v_1 \\
g \rightarrow d &\quad& \mathrm{if}\quad \theta  + v < v_0 
\end{eqnarray*}
} \right. \end{displaymath}

Ice volume $v$, expressed in sea-level equivalent, responds to the following equation:

\begin{eqnarray*}
\ddt{v} &=& - a_\Pi \Pi^\star  - a_{\tilde\Pi} * \tilde\Pi^\star - a_O \tilde O + 
\left\{
\begin{array}{ll}
a_d - v / \tau & \quad \textrm{ if state is $d$} \\
a_g            & \quad \textrm{ if state is $g$,} \\
\end{array}
\right.
\end{eqnarray*}
with the following parameter values:  \def\mka{\ensuremath{\mathrm{m/\ka}}}
$ a_\Pi = 1.456 \mka$, $ a_{\tilde \Pi} = 0.387 \mka$, $ a_{O}   = 1.137 \mka$, $ a_g   = 0.978$ \mka, $ a_d   = -0.747 \mka$, $ \tau = 0.834 \ka$, $ k_\Pi = 14.635$m, $ k_{\tilde\Pi} = 2.281$m, $ k_{O} = 23.5162$m, $v_0=122.918$m and $v_1=3.1031$m, assuming that one time unit$=10$ka. This parameter set is the one originally presented by \citeauthor{Parrenin12ab} in Climate of the Past Discussion (which differs from the final version in Climate of the Past), except that $k_{O}$ is $18.5162 \mathrm{m}$ in the original paper. This modification was needed to reproduce the exact sequence of terminations shown by the authors. Subtle details, such as the numerical scheme or the choice of the astronomical solution might explain the difference.

All codes and scripts are available from GitHub at 
\texttt{https://github.com/mcrucifix}.

\begin{acknowledgements}
Thanks are due to Peter Ditlevsen (Niels Bohr Institute, Copenhague), Fr\'ed\'eric Parrenin (Laboratoire de Glaciologie et de G\'eophysique, Grenoble), Bernard De Sae\-deleer, Ilya Ermakov and Guillaume Lenoir (Universit\'e catholique de Louvain) for comments on an earlier version of this manuscript. Thanks also to the numerous benevolent developers involved in the R, numpy and matplotlib projects, without which this research would have taken far more time. MC is research associate with the Belgian National Fund of Scientific Research. This research is a contribution to the ITOP project, ERC-StG grant 239604. 
\end{acknowledgements}

\bibliographystyle{copernicus}
\bibliography{/Users/crucifix/Documents/BibDesk.bib}

\end{document}